\documentclass[fleqn,usenatbib,useAMS]{mnras}

\usepackage{graphicx}	% Including figure files
\usepackage{amsmath}	% Advanced maths commands
\usepackage{amssymb}	% Extra maths symbols
\usepackage{multicol}        % Multi-column entries in tables
\usepackage{bm}		% Bold maths symbols, including upright Greek
\usepackage{pdflscape}	% Landscape pages

\usepackage[T1]{fontenc}
\usepackage{ae,aecompl}

\usepackage{newtxtext,newtxmath}
\title[GRB Jet Opening Angle]{The Evolution of Gamma-ray Burst Jet Opening Angle through Cosmic Time.}

\author[]{Nicole Lloyd-Ronning$^{1,2}$\thanks{Contact e-mail: \href{mailto:}{lloyd-ronning@lanl.gov}}, Valeria U. Hurtado$^{1}$, Aycin Aykutalp$^{3}$,
\newauthor Jarrett Johnson$^{3}$,  Chiara Ceccobello$^{4}$
\\
$^{1}$CCS-2, Los Alamos National Lab, Los Alamos, NM 87544\\
$^{2}$Department of Science and Engineering, University of New Mexico, Los Alamos, 87544\\
$^{3}$ XTD, Los Alamos National Lab, Los Alamos, NM 87544\\
$^{4}$ Chalmers University of Technology, Onsala Space Observatory, 439 92 Onsala, Sweden}

\date{Last updated 2015 May 22; in original form 2013 September 5}

\pubyear{2019}

\begin{document}
\label{firstpage}
\pagerange{\pageref{firstpage}--\pageref{lastpage}}
\maketitle

\begin{abstract}
  Jet opening angles of long gamma-ray bursts (lGRBs) appear to evolve in cosmic time, with lGRBs at higher redshifts being on average more narrowly beamed than those at lower redshifts.  We examine the nature of this anti-correlation in the context of collimation by the progenitor stellar envelope. First, we show that the data indicate a strong correlation between gamma-ray luminosity and jet opening angle, and suggest this is a natural selection effect - only the most luminous GRBs are able to successfully launch jets with large opening angles. Then, by considering progenitor properties expected to evolve through cosmic time, we show that denser stars lead to more collimated jets; we argue that the apparent anti-correlation between opening angle and redshift can be accounted for if lGRB massive star progenitors at high redshifts have higher average density compared to those at lower redshifts. This may be viable for an evolving IMF - under the assumption that average density scales directly with mass, this relationship is consistent with the form of the IMF mass evolution suggested in the literature.  The jet angle-redshift anti-correlation may also be explained if the lGRB progenitor population is dominated by massive stars at high redshift, while lower redshift lGRBs allow for a greater diversity of progenitor systems (that may fail to collimate the jet as acutely).  Overall, however, we find both the jet angle-redshift anti-correlation and jet angle-luminosity correlation are consistent with the conditions of jet launch through, and collimation by, the envelope of a massive star progenitor.  
  
\end{abstract}


%%%%%%%%%%%%%%%%%%%%%%%%%%%%%%%%%%%%%%%%%%%%%%%%%%

%%%%%%%%%%%%%%%%% BODY OF PAPER %%%%%%%%%%%%%%%%%%

% The MNRAS class isn't designed to include a table of contents, but for this document one is useful.
% I therefore have to do some kludging to make it work without masses of blank space.

\section{Introduction}

 Although we have learned much about gamma-ray bursts (GRBs) over the last 20 years, there are still a number of open questions related to the nature of their underlying progenitor systems. It is well established that they are associated with the deaths of massive stars and/or merging binary systems (for reviews summarizing the arguments and evidence for this, see \cite{pir04, ZM04, Mesz06, LRR07, GRRF09,Berg14,DAvanz15,Lev16}). However, it is also clear that very special conditions are required to successfully launch a GRB jet.  These conditions distill down to the inner engine having enough angular momentum and power to launch a relativistic jet that can propagate through a surrounding hydrogen-stripped envelope. 
  
  There are many proposed systems capable of producing a GRB and in reality, probably several (or all) of these systems contribute to the total GRB population. \cite{Lev16} summarize different progenitor systems for lGRBs, including various single star and binary formation channels, the rates of these different formation channels, and other important considerations (e.g. host galaxy properties, etc.).  Signatures of specific progenitor systems, like locations in their host galaxies \citep{BKD02,Fong2010, Fong2013, Ly17}, or coincident emission associated with a supernova \citep{Gal98,Hjorth03,WB06,HB12}, kilonova \citep{Metz10,TL13,Tro18}, or the presence of gravitational wave emission \citep{Ab17} can potentially help distinguish among progenitor systems for both long and short GRBs.  In addition, there are many correlations present among observed and fitted GRB variables (e.g. see the recent review by \cite{Dai18}), which may help to elucidate the underlying progenitor.  
 
 In this paper, we consider how different progenitor systems connect to GRB observables over cosmic time.  We are motivated by the results of \cite{LR19} who found that certain intrinsic long gamma-ray burst (lGRB) properties appear to evolve with redshift, even when accounting for Malmquist-type biases and selection effects in the observed data. This and previous studies \citep{LRFRR02,WG03,Yon04,KL06,Yu15,PKK15,Deng16,TSv17,Xue19} have shown that isotropic energy and luminosity evolve as a function of redshift, with lGRBs being brighter at higher redshifts even when accounting for selection effects that favor detecting more luminous bursts at high redshifts.  However, in the hundred or so bursts where jet opening angle estimates are available and for which one can compute beaming angle corrected (i.e. actual emitted) gamma-ray energy and luminosity, \cite{LR19} found these variables (i.e. gamma-ray luminosity and emitted energy) are {\em not} correlated with redshift.  This suggests  that jet opening angle {\em is}, and indeed they found a significant  anti-correlation between jet opening angle and redshift.  This relationship between jet angle and redshift was suggested in \cite{LRFRR02} (e.g. see their section 5.1.2), and observational evidence for this anti-correlation has also been put forth by \cite{Yon05, Lu12} and \cite{Las14,Las18,Las18b}.  
 %In addition, the data indicate that prompt gamma-ray durations also evolve as a function of redshift, with higher redshift lGRBs having a shorter intrinsic prompt gamma-ray duration. We would like to understand the nature of this evolution and how it relates to  potential lGRB progenitors.  \\

 If this apparent anti-correlation - with higher redshift lGRBs more narrowly beamed than lower redshift lGRBS - is true, its underlying cause is not clear.  Beaming angle evolution over cosmic time could be a reflection of the evolution of any number of properties or processes that affect the lGRB jet  - for example, the average stellar density profile, the spin and magnetic field geometry of the central engine, etc.  This correlation can also be affected by the degree of sideways spreading of the jet when it breaks free of the star, as well as inherent and observational selection effects. \\

 Our goal in this paper is to understand the potential beaming angle evolution in the context of collimation by the stellar envelope, and how it relates to lGRB progenitor properties.  Our paper is organized as follows.  In \S 2, we summarize the general properties of different GRB progenitor models.  In \S 3, we examine the luminosity requirements both for launching a successful jet and for collimation by the stellar envelope. We show there is a strong correlation between emitted gamma-ray luminosity and jet opening angle, and suggest this is a natural selection effect where only the more luminous lGRBs can launch wider opening angle jets. In \S 4, we discuss collimation of the jet by a massive star envelope and how evolution of certain intrinsic properties such as mass and metallicity can lead to the evolution of the jet opening angle. In particular, we show that the beaming angle-redshift anti-correlation can be explained if high redshift lGRB progenitors are denser compared to those at lower redshift, which may be expected for an evolving IMF. In \S 5, we present a summary and our conclusions.  

\section{Data Sample and Selection Effects}

 Our data are taken from \cite{Wang2019}, who compiled publicly available observations and fitted GRB parameters for $6289$ gamma-ray bursts from 1991 to 2016. This data sample contains all previously published jet opening angle estimates (each entry in their electronically available table gives the reference where they obtained the data point), with a total of 137 lGRBs for which these data are available.  These jet opening angles are determined from a break (specifically, a steepening) at a given point in time in the afterglow (X-ray, optical and/or radio) light curve; the time of the break reflects when the forward blast wave has decelerated to a point at which the relativistic beaming of the radiation is comparable to the physical jet opening angle \citep{Rhoads99}.  Although this is a well-established and widely accepted method to get a reasonable estimate of the GRB jet opening angle, these methods rely on assumptions about the GRB outflow and jet structure, as well as its external density profile. We discuss the physical nature and potential uncertainties in the jet opening angle estimates further in \S 5 below. 

\subsection{Statistical Analysis of the Correlation}
 In \cite{LR19}, we found an anti-correlation between beaming angle and redshift with that can be parameterized as $\theta_{j} \propto (1+z)^{-0.75 \pm 0.25}$.  In that paper, we use well-established non-parametric techniques \citep{LB71,EP92} to account for Malmquist biases in the detection of high redshift GRBs.  To estimate the significance of the correlation, we use additional non-parametric rank-order correlation tests (such as Kendell's $\tau$ and Spearman rank; for a nice explanation of these tests, see, e.g., \cite{Press07}).  These types of statistical tests are useful because there are no assumptions about the underlying distributions of the variables being evaluated.  However, they have other drawbacks - in particular, they do not easily account for measurement error (see, e.g., \cite{Kit18}).  There are techniques developed to estimate correlations accounting for measurement error \citep{Kell07}, and these are particularly powerful when the error dominates the scatter in the data.  However, these techniques again rely on making some assumptions (however reasonable) about the underlying distributions of the variables and their errors.  In our case, because the scatter in the data is larger than the data points' error bars, we do not expect that the presence of measurement error will significantly affect the correlation.  To test this statement, however, we performed a series of simulations in which we drew each data point from a uniform distribution within in its $1 \sigma$ error bars.  In all cases, we still found an anti-correlation between $\theta_{j}$ and $(1+z)$ to high statistical significance. \\
  
  For the purposes of this study, the results of the rank correlation tests provide a fairly robust method of estimating the probability of a correlation in the data. However, the presence of observational selection effects can in some cases lead to false (statistically significant) correlations in the data.  We discuss this important point in the following section. \\

\begin{figure}
  \includegraphics[width=3.5in]{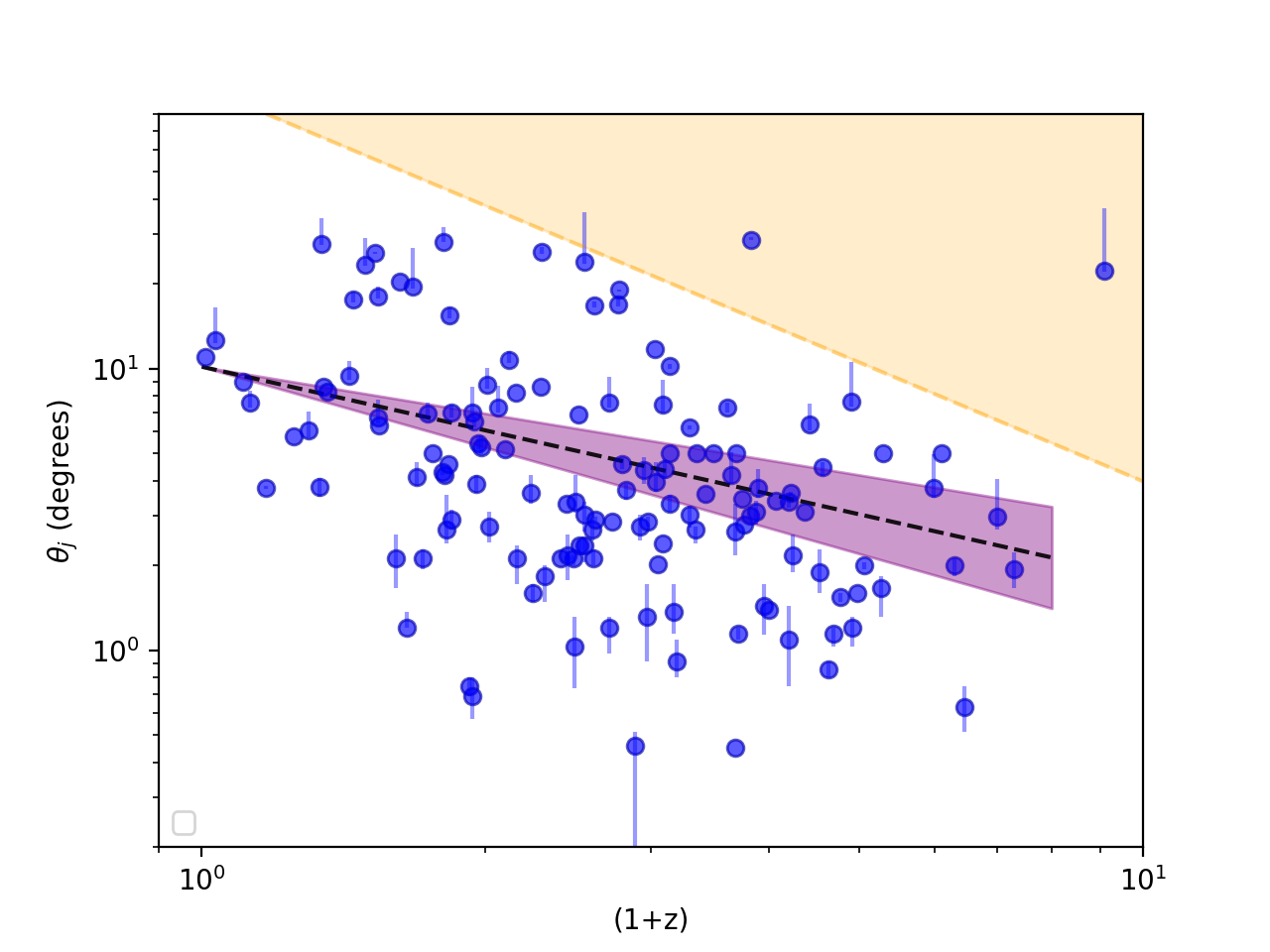} 
  \caption{Jet opening angle vs redshift. The dashed black line and purple region show the best fit to the data, $\theta_{j} \propto (1+z)^{-0.75 \pm 0.25}$.  The orange shaded area indicates a region of possible selection bias.}
  \label{fig:thetajztrunc}
  \end{figure}     

\subsection{The Role of Selection Effects}

Before discussing the potential physical nature of jet opening angle evolution, it is important to consider whether selection effects may be playing a role in producing the apparent relationship. For example, \cite{Lu12} found a similar statistically significant correlation between jet opening angle and redshift in their sample of 77 lGRBS, 
with $\theta_{j} \propto (1+z)^{-0.94 \pm 0.19}$.  They argue, however, that the correlation can be explained by observational biases.  Their analysis relies on having to assume specific functional forms for their underlying variables, including the lGRB rate density, luminosity function, jet beaming angle distribution, and additional assumptions about the detector response rate.  Importantly, they have assumed that the lGRB rate density follows the star formation rate, a conjecture that has been called into question by a number of studies, based on lGRB observations (e.g. \cite{LRFRR02, Kist08, Yuk08, Kist09, PKK15, LR19}).  Because of the degeneracy among the physical variables and the lack of knowledge of their true underlying distributions, it is difficult to draw definitive conclusions about the role of selection effects through this approach.  As we discuss below, non-parametric techniques developed to deal with fairly well-defined selection criteria can be a useful tool to determining the true correlation between variables \citep{LB71, EP92}.  To do so, however, we need to get a handle on what these selection effects are. \\

As discussed in the introduction, many of the issues related to Malmquist-like selection effects in this sample have been addressed in \cite{LR19}. However, it is important to ask how missing a fraction of low(er) luminosity GRBs at high redshift due to detector sensitivity limits might affect our results.  Such a selection effect would lead to a corresponding lack of lower luminosity, smaller jet angle data points (because jet angle is correlated with luminosity; see \cite{Lu12} and \S 4 below).  Therefore, adding these data points to our sample - if indeed they are there and we are simply not seeing them - would strengthen the anti-correlation between jet opening angle and redshift. That is, because luminosity is correlated with jet opening angle, missed low luminosity GRBs at high redshift would have on average smaller opening angles.  Adding such GRBs to our sample would then place more small opening-angle GRBs at high redshift, strengthening our correlation. \\
%Additionally, it has been suggested that GRBs at high redshift are inherently more luminous, allowing for larger opening angles (see \S 4.1 below); accounting for this would strengthen the anti-correlation between beaming angle and redshift. Therefore, we do not believe this observational selection effect is playing a large role in producing the $\theta_{j}-(1+z)$ correlation.

However, another important possibility is that we are missing a fraction of larger opening angle GRBs at high redshift, because the jet break occurs too late in the afterglow light curve to be detected (i.e. the afterglow light curve has faded below detector limits before the jet break can be detected).  There are a couple of things that mitigate this effect, however.  First, lGRBs are not standard candles so the intrinsic scatter in their afterglow luminosities helps soften this effect to some extent.  Additionally, lGRBs with larger opening angles should have larger average luminosities, and are less likely to be missed by the detector (relative to low luminosity lGRBs); therefore we suggest this effect may not be playing a significant role in artificially producing a $\theta_{j}-(1+z)$ anti-correlation.\\
%In addition, and related to the points in the previous paragraph, if lGRBs really are more luminous at high redshift, we expect that to be true for their afterglows as well since there is a well-established relationship between the gamma-ray and afterglow luminosities (CITE).  

To examine the severity and effect of this latter potential bias, we have produced a strong truncation in the $\theta_{j}-(1+z)$ plane, in which larger jet opening angles at high redshift are missed due to observational selection effects.  This is shown by the orange shaded region in Figure~\ref{fig:thetajztrunc} (note this region excludes a couple of observed lGRBs, and so is potentially more severe than necessary).  Applying this truncation to our sample and performing the rank statistics discussed in \cite{LB71,EP92}, we still find a significant anti-correlation between $\theta_{j}$ and $(1+z)$.  Again, we favor the non-parameteric approach in determining the significance of the correlation between variables because we don't need to make assumptions about the parent population distributions, and they have been shown to reveal true underlying correlations very well.  For example, the appendix of \cite{LPM00} demonstrates how powerful these methods can be in recovering the underlying distributions of relevant variables (and the correlation between them) in the presence of truncation.  In their Figures 7 and 8, they show simulations of two important cases: 1) when the truncation produces an artificial correlation, and 2) when the truncation removes a true, underlying correlation.  In both cases, applying the methods of \cite{LB71, EP92} reproduces the true, underlying distributions of, and correlations among, the variables at hand.  Other methods to incorporate selection effects based on a Bayes approach are described in \cite{MFG19} (and although powerful, these methods do rely on making some assumptions about the underlying distributions of the variables). \\

 There are a variety of approaches one can employ to deal with complicated observational selection effects. In \cite{LR19} (where we initially report this correlation) and in this paper, we have chosen to use well-established non-parametric rank-order techniques because - again - they don't require assumptions about the underlying distributions of the variables being analyzed.  The complexity of this issue, however, may leave some room to question the presence of a physical anti-correlation between $\theta_{j}$ and $(1+z)$. We emphasize that the purpose of this paper is to explore a possible physical interpretation behind this potential correlation. In other words, as we show below, we {\em predict} such an anti-correlation between lGRB beaming angle and redshift should exist if stars in the early universe are more massive (on average) and/or have lower metallicity than stars at lower redshift. \\

\section{Progenitor Models}
   As mentioned in the introduction, there are several general requirements that appear necessary to produce a successful relativistic GRB jet. These include: {\bf 1)} enough mass and angular momentum in the system to sustain an accretion disk and launch a jet,   {\bf 2)}  no hydrogen envelope; this is based on both theory (the requirement that the jet is able to breakout from the system's envelope) and observations of Type Ic supernova associated with lGRBs, and {\bf 3)}  significant magnetic flux (along with angular momentum) to launch a jet (assuming a magnetically launched jet as in the Blandford-Znajek framework \citep{BZ77}.  See, e.g., \cite{BK08,KB09, BK10, LR19b} for a consideration of these conditions for lGRB systems). 
   %Note that magnetic torques also tend to produce rigid rotation of the star.  However, one can potentially get around this by growing the B-field through some instability like an MRI.
   We discuss these requirements briefly and generally in the context of several of lGRB progenitor systems.
  
\subsection{Single Massive Star Progenitors}
A number of studies have shown the viability of a massive star progenitor for lGRBs from a theoretical point of view (e.g., \cite{Woos93, MW99, WH06, KNJ08a, KNJ08b}, and recently \cite{OA19}).  In addition, there is strong observational evidence, both through supernova associations (see, e.g. \cite{HB12}) and locations in star forming regions in their host galaxies \citep{BKD02,Ly17} that lGRBs are associated with the deaths of massive stars.  [We note there are two examples (GRB060505 and GRB060614, which lie in the in-between ground of lGRB and sGRB in terms of duration) that do {\em not } have associated SNe to deep limits \citep{DV06, Fyn06, GY06, Gehr06}].  

How a collapsing star loses its hydrogen envelope and retains the angular momentum to sustain a disk and launch a jet is a complicated question.  Stars with higher metallicity are particularly problematic because the associated mass loss carries angular momentum. Lower metallicity stars may be more viable candidates, since mass loss goes roughly as $Z^{0.7-0.8}$, where $Z$ is metallicity \citep{Vink01,VdK05}.  If one has a low metallicity star, therefore, one might expect the necessary angular momentum can be retained, and indeed this is one of the motivations behind considering Population III stars as GRB progenitors \citep{BL06,SI11,Yoon12}. However, other factors such as torques from magnetic fields and/or other coupling between the elemental layers of stars can cause a loss of the necessary angular momentum. Chemically homogeneous evolution helps mitigate the latter effect. This is addressed in \cite{WH06}, who assert about 1\% of stars can achieve the conditions needed to retain enough angular momentum to launch a GRB (keep in mind that their models are one dimensional and asymmetric mass loss may help, allowing for less angular momentum loss).   However, this process needs to occur during the early stages of a star and - again - is more efficient for lower metallicity stars (further discussion of these issues can also be found in section 5.2 of \cite{Lev16}).

 %The requirement of losing the hydrogen envelope is also a problem because single massive stars lose their H envelope through mass loss which takes angular momentum away from the star.  Of the single massive star scenario, both higher mass, lower metallicity PopIII stars are viable \citep{SI11,Yoon12} (which we'd expect at higher redshifts), or more metal rich, lower mass PopII stars. {\em Aycin and Jarrett comments here?}

 Although massive stars can span a range of mass and metallicities, and can meet the energy and timescale requirements for a lGRB,  the difficulties associated with retaining enough angular momentum are one of the motivations for considering binary formation channels for lGRBs.

%    {\em Side note: If a star forms a disk before collapse (e.g. see pl 637 of \cite{WH06}, could you have an outer disk and then an inner disk and perhaps that outer disk could serve as the mass/energy source to re-energize the GRB and produce a late time flare?}

\subsection{Binary Formation Channels}
Binary systems are thought to make up at least half of the massive stars \citep{Kob14}.  As with single massive star progenitors, binary lGRB progenitors need to strip the hydrogen envelope and retain enough mass and angular momentum to allow for  GRB jet launch.  The advantage to binary progenitors is that the conditions of hydrogen envelope stripping and high angular momentum are in principle readily met, due to the interaction with the companion star \citep{BBR02,FH05,KA17,Dav07,BK10,deMink13, Lev16, CSE19}.
%This will occur for a particular separation of the binary - these issues are discussed in \cite{Dav07}.  See the constraints on separation in Equation 2 and Figure~\ref{fig:thetaz}. 
\\

\noindent Viable binary progenitors for lGRBs include:\\ 
%    \begin{itemize}
     \noindent {\em -- Helium Mergers} \citep{FW98,ZF01,Fry13}.  In this model, a compact object merges with the helium core of an evolved companion.  In the process, the hydrogen envelope is ejected and the compact object is spun up so that the conditions for launching an lGRB jet are met.   This model can achieve the necessary energetics and timescales to be a viable lGRB progenitor.\\ 
     %{\bf Is there a distinction from the secondary going supernova and making a GRB (while in orbit with a black hole that gave it enough angular momentum vs. the case where the compact object (e.g. BH) spiraled into the core of the He star and then the star went SNe? Maybe no matter - maybe all about how to get enough angular momentum and create the necessary BH-disk system!}}
     \noindent {\em -- CO core mergers}.  A related model is the so-called binary-driven hypernova involving accretion induced collapse (for a recent summary of this model see \citep{RRW19}). In this scenario, a carbon-oxygen core undergoes a supernova which causes rapid accretion onto a neutron star companion. Similar to the He-merger model, it can achieve the conditions required to launch a relativistic GRB jet.\\ 
     \noindent {\em -- White Dwarf-Black Hole/NS binaries} \citep{King07}.  In this model, a white dwarf merges with a black hole or neutron star companion and can produce a long gamma-ray burst, in some cases without an accompanying supernova.\\ 
     \noindent {\em -- Micro-TDEs} \citep{Per16}.  In this model a neutron star or black hole tidally disrupts a star, leading to a debris disk around the compact object that can launch a relativistic jet. This model was initially proposed to explain ultra-long GRBs, but may contribute to the standard lGRB population.\\
     
     Other binary systems not considered here (e.g. \cite{Cant07,Call19}) may also contribute the the lGRB population.  Because of the vast array of binary systems, their formation channels and subsequent evolution, they may span a range of masses/energies, metallicities, angular momenta, and have a range of different ambient environments. We note that some simulations (e.g. \cite{Yoon10}) of binary progenitors for lGRBs - particularly those able to produce Type Ic SNe -  find in fact not enough angular momentum is retained to launch an lGRB. As with massive star progenitors, however, lower metallicity conditions in these systems may help \citep{KA17}, as well as a stage of chemically homogeneous evolution (e.g. see the recent discussion in \cite{CSE19}).

\subsection{Expected Cosmological Evolution Of GRB Intrinsic and Environmental Properties}
 The primary properties of lGRB progenitors expected to evolve through cosmic time are stellar mass and metallicity (which are of course themselves related). Metallicity evolves with redshift (i.e. \cite{Vink01,VdK05}) as stars in the early universe have not had the time to synthesize a large amount of metals.  A star's metallicity can affect its mass, spin, and stellar density profile.  For example, the distribution of zero-age main sequence stellar masses is expected to evolve over cosmic time largely as a result of lower metallicity conditions allowing more massive stars to form \citep{Kroupa19}. Evidence for the evolution of this initial mass function (IMF) has been suggested by a number of authors (e.g. \cite{vD07,Dave08, WD11, Mark12} and more recently \cite{Leja19,Chrus20}).  Overall stellar mass can affect the lGRB energy budget and - as we discuss below - the stellar density profile, where more massive stars may serve to collimate a jet more effectively. 
 
\section{Luminosity Requirements for Successful Jet Launch and Collimation}
\begin{figure*}
  \includegraphics[width=6.5in]{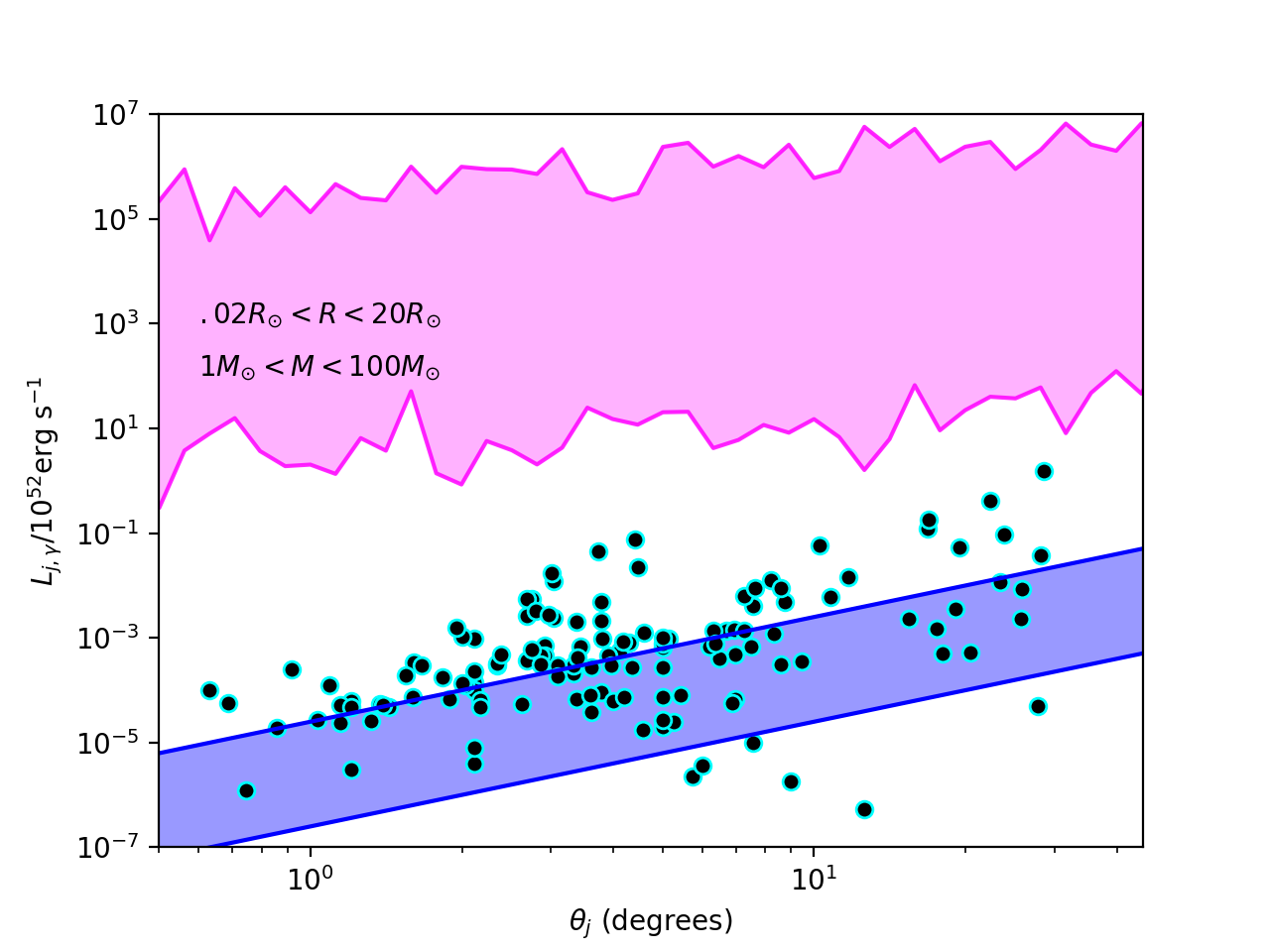} 
  \caption{Gamma-ray luminosity vs. jet opening angle. The upper magenta band shows the upper limit to the jet luminosity required to ensure a collimated jet \citep{Matz03,Brom11} for extreme ranges of mass and progenitor stellar radii.  The fluctuations are due to a random selection of stellar density profile indices ranging from $-3.0 < \alpha < -2.1$, where $\rho \propto r^{-\alpha}$.  The lower blue band is the luminosity needed to launch a jet \citep{Aloy18} given a jet injection radius of $h_{o} = 2x10^{9}$ cm and an ambient pressure of $1.8x10^{23}$ erg cm$^{-3}$, representative of a $35 M_{\odot}$ zero-age main sequence star with enough angular momentum to launch a GRB \citep{WH06,Aloy18}.  The blue band spans  a gamma-ray efficiency (converting jet power to radiation) of $0.1 \%$ to $10\%$.} %(Alternatively, you get a similar band for pressure spanning $1.8x10^{21}<p_{a}<1.8x10^{23}$ erg cm$^{-3}$).}
  \label{fig:lumtheta}
  \end{figure*}     
  
 It takes a certain amount of power in order for a jet to borough through a stellar envelope (although there may exist a pre-jet that evacuates the polar region, easing this requirement, e.g. \cite{Burr07}). A key quantity that needs to be considered is the ratio of the jet energy density to ambient medium energy density, given by \citep{Matz03,Brom11,HKI19}:
   \begin{equation}
       \widetilde{L} = \frac{\rho_{j}\eta_{j}\Gamma_{j}^{2}}{\rho_{a}\eta_{a}\Gamma_{a}^{2}}
   \end{equation}
   \noindent where $\rho$ denotes density, $\eta$ denotes the specific enthalpy and $\Gamma$ denotes the Lorentz factor.  The subscript $j$ refers to the jet while $a$ refers to the ambient medium.  For a static ambient medium $\Gamma_{a} = 1$ and $\eta_{a} = 1$.
   
  Given this ratio and the requirement that - as discussed in \citep{Aloy18} - the outflow must be supersonic with respect to the external medium, there is a requirement on the minimum power a jet must have to launch successfully \citep{Aloy18}:
  
  \begin{equation}
       L_{j} \gtrsim 10^{49} {\rm erg s^{-1}} \left(\frac{h}{2x10^{9}{\rm cm}}\right)\left(\frac{\theta_{j}}{2^{\circ}}\right) \left(\frac{p_{a}}{1.8x10^{22} {\rm erg cm^{-3}}}\right)
   \end{equation}
   
\noindent where $h$ is the height of the jet, $\theta_{j}$ is the jet opening angle, and $p_{a}$ is the pressure in the ambient medium.

   As shown in \cite{Matz03, Brom11}, there is also an upper limit to the luminosity to ensure the jet is collimated in the stellar envelope.
   The condition for collimation is given by \citep{Matz03, Brom11}:
   \begin{equation}
       \widetilde{L} < \theta_{o}^{-4/3}
   \end{equation}
   
  \noindent where $\theta_{o}$ is the initial jet opening angle.
  For $\eta_{a}, \Gamma_{a} = 1$, this condition becomes
  \begin{equation}
       \widetilde{L} = \frac{\rho_{j}\eta_{j}\Gamma_{j}^{2}}{\rho_{a}} \sim \frac{L_{j}}{\Sigma_{j}\rho_{a}c^{3}} < \theta_{o}^{-4/3}
   \end{equation} 
   \noindent where $\Sigma$ is the cross sectional area of the jet.  Clearly a key parameter is the stellar density profile, which we expect to go roughly as $\rho_{a} \sim r^{-3}$ \citep{MM99}, and this is indeed borne out by numerical simulations of massive stars \cite{WH06, MA09}.  Chemically homogeneous mixing can affect this profile, however \citep{WH06}, and it has been shown that, more generally, massive star density profiles go as $\rho \propto r^{-n}$, with $n$ between about 2 and 3 \citep{MM99} out to the radius of the helium envelope, with a sharp drop off thereafter.  The steepness of the drop off is strongly dependent on the mass loss/metallicity, with low metallicity stars dropping off more drastically \citep{WH06, MA09, SI11}.  Following \cite{Brom11}, we write the ambient density as $\rho_{a} = \overline{\rho} (h/R)^{-\alpha}$; the parameter $\overline{\rho} = ((3-\alpha)/4\pi)M R^{-3}$ is the average stellar density, where $M$ is the mass of the star and $R$ is the stellar radius.

  Then, the condition from equation 4 above can be written as \citep{Brom11}:
  \begin{equation}
      L_{j} < 10^{54}\rm{erg s^{-1}}\left(\frac{R}{R_{\odot}}\right)^{-1}\left(\frac{h}{R}\right)^{2-\alpha} \left(\frac{\theta_{o}}{10^{\circ}}\right)^{2/3} \left(\frac{M}{10M_{\odot}}\right)
  \end{equation}
  
  These luminosity conditions for jet launch and collimation (i.e. equations 2 and 5) are shown in Figure~\ref{fig:lumtheta} along with the data (gamma-ray luminosity, $L_{\gamma}$, versus measured jet opening angle; black-cyan dots).  The upper pink band shows the collimation condition (equation 5) - the upper limit to the luminosity for a mass range and radius range of $0.02R_{\odot} < R < 20 R_{\odot}$ and $M_{\odot} < M < 100 M_{\odot}$.  The fluctuations reflect a random selection of stellar density profile indices ranging from $-3.0 < \alpha < -2.1$.  The lower blue band shows the minimum luminosity (equation 2) needed to launch a jet \citep{Aloy18} given a jet injection radius of $R_{j} = 2x10^{9} {\rm cm}$ and an ambient pressure of $1.8x10^{23}$ erg cm$^{-3}$, representative of a $35 M_{\odot}$ zero-age main sequence star \citep{WH06,Aloy18}.  The band spans a gamma-ray efficiency (converting jet power to radiation) from $0.1 \%$ to $10\%$. 
  %(Alternatively, you get a similar band for a range of pressures spanning $1.8x10^{21}<p_{a}<1.8x10^{23}$ erg cm$^{-3}$). 
  
  \subsection{On the Observed Correlation between Luminosity and Jet Opening angle}
    There is a strong correlation between gamma-ray luminosity and measured jet opening angle.  The best fit line to this correlation is $L_{\gamma} \propto \theta_{j}^{1.5 \pm 0.2}$.  Note that the data points that fall below the minimum jet luminosity requirement (blue band in Figure~\ref{fig:lumtheta}) do not necessarily violate the minimum luminosity requirement - they may simply be indicating a lower efficiency of conversion from jet power to gamma-ray luminosity than we have assumed ($1 \%$). That is, the emitted luminosity is only a fraction of the jet energy, $L_{\gamma} = \epsilon L_{j}$, and this fraction may lower than $1 \%$ for some lGRBs. It is also possible, however, that these GRB progenitors have a smaller envelope pressure than we have assumed ($1.8x10^{21}<p_{a}<1.8x10^{23}$ erg cm$^{-3}$).  Either one of these possibilities, however, may say something about the progenitor and/or its environment.\\

  Remembering the pressure in the jet scales as $L_{j}/\theta_{j}^{2}$ and the minimum luminosity to launch a jet scales as $\theta_{j}$, we can speculate that this correlation arises naturally in a massive star progenitor scenario.  Successfully launched jets may sit at a base value of the pressure balance condition (otherwise they may be swallowed by their cocoons), but necessarily satisfying the minimum luminosity condition.  The location of the data points in Figure~\ref{fig:lumtheta} seems to suggest the jet launch condition plays a larger role in producing this correlation. Ultimately, then, this correlation reflects the successful jet launch condition - only jets with high luminosities can have wide opening angles.  This correlation might also be explained if those lGRBs with higher luminosity have higher internal energy in the jet and undergo more sideways expansion such that a wider jet is measured at the time of the afterglow ``jet break'' (we discuss this further below).

 %Note that luminosity-variability correlations would argue the opposite - higher luminosity would have smaller opening angles if variability was inversely correlated with jet opening angle (e.g. see \cite{SG02}). \\
  
 %The correlation between gamma-ray luminosity and jet opening angle seems to run contrary to studies that have suggested the variability of an lGRB is related to the jet opening angle (e.g. \cite{SG02,KRM02}), with more variable lGRBs having narrower jet opening angles.  Variability has been shown to be correlated with gamma-ray luminosity \citep{FRR00,Reich01} and therefore we would expect an anti-correlation between jet opening angle and luminosity in this case.  Because we see the opposite trend, we suggest that the prompt emission light curve variability of an lGRB is not strong related to the opening angle of the jet (RR & L-R. paper?)
  
  % Kobayashi, Ryde and MacFadyen.

 %\section{Energy Evolution}
 
 \section{Beaming Angle Collimation and Evolution}
 
If the anti-correlation between beaming angle and redshift is indeed physical, we would like to understand what determines the observed jet opening angle and how it could evolve through cosmic time - i.e. its relationship to fundamental progenitor properties.  Besides the observational selection effects discussed in \S 2, there are a number of issues to consider when attempting to understand the nature of the jet opening angle:  
 
     \noindent -- How does the jet launching mechanism affect the initial jet opening angle?  For example, in a Blandford-Znajek framework, one should consider how the spin of the black hole and magnetic field strength and geometry serve to constrain the angle of the outflow.  Naively, for example, one might expect a higher black hole spin could serve to wind the magnetic field more tightly and lead to a more collimated jet.  If lGRB progenitors in the early universe produce on average more highly spinning black holes or have some property of their magnetic fields that leads to more collimated jets, this could play a role in the beaming angle evolution.\\
     \noindent -- How the stellar density profile and jet cocoon \citep{RR02} structure will serve to collimate the jet. \\
     \noindent -- The sideways expansion of the jet once it breaks out of the star.\\
     \noindent -- How the angular structure of the jet plays a role in estimates of the jet opening angle.\\
  
  The first issue is a complicated question and is perhaps best investigated with detailed GRMHD simulations of black hole-disk-jet systems (Hurtado et al., in prep).  Our focus in this paper is on the second issue.  However, we briefly address the third and fourth issues below.\\

 \noindent{\em On Sideways Expansion}\\
 \indent It is important to consider what the inferred lGRB beaming angle actually reflects.  When the outflow velocity of the jet is decelerated to a point where relativistic beaming ($\sim 1/\Gamma$) is on the order of the physical jet opening angle $\theta_{j}$, photons are able to escape ``sideways'' and a steepening of the light curve will occur \citep{Rh97,Rhoads99}.  For a uniform jet, the temporal behavior of the jet Lorentz factor decelerating in a constant medium is $\Gamma(t) \propto (E/n)^{1/8} t^{-3/8}$, so that $\theta_{j} \propto (E/n)^{-1/8} t_{b}^{3/8}$, where $t_{b}$ is the time of the jet break. Therefore, a break in the afterglow light curve gives an estimate of the jet solid angle at the time of this observed jet break.  If the jet has undergone significant sideways expansion, then this opening angle is not an accurate reflection of the true opening angle of the jet as it emerged from the star/progenitor. \\

 Several numerical simulations have examined this problem and shown that the jet does in fact {\em not} undergo significant sideways expansion, so that it is not unreasonable to take the opening angle at breakout to be the asymptotic jet angle \citep{MA09, Tch10,Aloy18}.   In other words, in many cases the jet is ``ballistic'' and the estimated jet opening angle can be considered a decent estimate of the true jet opening angle.  
%See, e.g. the last paragraph of \S 3 of \cite{MA09} who argue that sideways expansion is strongly suppressed. 
%In contrast, \cite{ZWM03}, who examine a jet from a  a $15 M_{\odot}$ Helium star with $10\%$ solar metallicity, and with an initial surface rotation of 10\% Keplerian, found the jet   Nonetheless, they find on average a final jet opening angle of $\theta_{o} + 1/\Gamma_{o}$ from their simulations; for high enough initial Lorentz factors $\Gamma_{o}$, therefore, the final jet opening angle is a good estimate of the initial opening angle.
%Bottom line, for the flow to expand sideways after breakout, we need internal energy in the jet to do so - $h_{j}\rho_{j} \approx 4p_{j}$ (CITATION) and $\Gamma < 1/\theta_{j}$.  \\  
 
  However, \cite{ZWM03} find a jet can undergo some amount of sideways expansion when it emerges from the stellar envelope if it has high enough internal energy.  In addition, \cite{MLB07} run FLASH simulations of GRB jets propagating through stars. In particular, they consider a star of low metallicity ($1\%$ solar), with an initial mass of $16 M_{\odot}$, a final mass of $13.95 M_{\odot}$ and a final radius of $4x10^{10}cm$.  They use a stellar density profile of $\rho \propto r^{-2.5}$ and an equation of state of $p = p_{o}\rho^{4/3}$ (i.e. ultrarelativistic), ensuring pressure in the star is small compared to $\rho c^{2}$.  They find the jet goes through three phases - a wide opening angle precursor phase, a shocked phase with a narrow opening angle (while in the stellar envelope, being collimated by the stellar envelope), and then an unshocked phase where the jet opening angle increases logarithmically in time (after break out).     
 Finally, \cite{Tch10} found, using GRBHMD simulations of magnetically launched jets that achieve the necessary (observational) constraint of $\theta_{j}\Gamma \sim 20$, that - except during an initial short acceleration phase - their jets show little sideways expansion after breakout. 
 
 In what follows, we assume the measured opening angle at the time of the break in the afterglow light curve is a decent approximation of the opening angle at breakout (or at least scales with the opening angle at breakout in a uniform way among all GRBs). \\ 
 
\noindent{\em On Angular Energy Distribution vs. Jet ``Opening Angle''}

 Another thing to keep in mind when considering jet ``opening angle'', is the angular energy distribution in the jet.   This point is addressed in detail in \cite{MA09}, who examine this distribution, $dE/d\Omega$, for different stellar envelope models (taken from \cite{WH06}).  Assuming the same mass, momentum and energy fluxes for all of their models, they find that the angular energy distribution decays more steeply for lower mass progenitors than high mass ones.  This is because, they argue, the average density of the progenitor grows approximately with mass, and as the mass (density) of the progenitor increases it leads to a slower jet propagation speed inside the star.  They suggest that jet slower speed allows for thicker, hotter cocoons to develop and therefore a broader range of $dE/d\Omega$.   In what follows below, we assume a relatively steep angular decay for the jet energy distribution so that the ``opening angle'' refers to the angle in which most of the jet energy is concentrated.  
 %For example, see figure 18 of \cite{MLB07} - their inner black line marks the jet boundary

 \subsection{Jet Collimation by the Stellar Envelope}
 
 There are a number of factors that come into play when considering whether a jet is collimated by its stellar envelope/cocoon system. These are discussed in detail in \cite{Matz03} and \cite{Brom11} among others.  As discussed in \S 3, these works address the necessary conditions for collimation as a jet propagates through a stellar envelope and forms a cocoon (which ultimately serves to collimate the jet, \cite{RR02}).     
Their analytic estimates consider mildly to non-relativistic jet heads - as \cite{Brom11} points out, if the head velocity exceeds a certain limit, the cocoon pressure will be too low to collimate it (as they show, this occurs when $L_{j}/(h^{2}\rho_{a}c^{3}) \approx \theta_{o}^{5/3}$).  This corresponds to a jet head Lorentz factor of $\Gamma_{h} \approx \theta_{o}^{-1/3}$, which is at most mildly relativistic.\\
   
 Numerical simulations that examine the extent of collimation can be found in \cite{ZWM03,Miz06, MLB07, Tch10}, all with similar general trends. \cite{ZWM03} find a jet with an initial half opening angle of 20 degrees propagating through a low metallicity massive star emerges with an opening angle of about 5 degrees.  \cite{Tch10}  suggest a relationship between the jet opening angle and stellar radius based on their GRMHD simulations of magnetically launched jets. In particular, they find the scaling $\theta_{j} \propto (1/R)^{0.22}$.  However, we caution that  this relationship could arise by design from the way they defined the wall of their jet as input (e.g. their equation 1).  Finally, we note that \cite{Nag14} and \cite{HKI19} have examined this problem in the case of a DNS merger scenario.  They also find significant jet collimation by the ejecta from the merger.  The latter in particular consider the ejecta velocity and its effect on collimation (see their Appendix C).
 These results suggest a general trend toward greater collimation, the longer the jet remains in the star (provided the envelope-cocoon system has enough pressure to collimate).  
  
 We would like to get a basic intuition about the extent of the jet collimation by the stellar envelope, and understand how it compares in the context of different progenitor scenarios. There are several ways to approach this but as a first step we can simply compare estimates of the energy densities in the cocoon and jet.  The cocoon pressure can be estimated by $p_{c} = \rho_{a}(\beta_{c} c)^{2}$, where $\beta_{c}$ is the velocity of the cocoon; the jet energy density is given by $L_{j}/(\Sigma_{j}c)$, where $\Sigma_{j} \sim \pi \theta_{j}^{2} h^{2}$ is the jet cross sectional area.   In that case, we find the jet opening angle scales as:
   \begin{equation}
   \theta_{j} \propto L_{j}^{1/2} \rho_{a}^{-1/2} h^{-1}. 
   \end{equation}
 
   Again, a key parameter in understanding the jet collimation is the stellar density profile.  Given that  $\rho_{a} = \Bar{\rho}(h/R)^{-\alpha}$ \citep{MM99}, where $2\lesssim \alpha \lesssim 3$, $\Bar{\rho}$ is the average stellar density, and considering the jet angle at breakout (i.e. $h = R$), we  find:
  \begin{equation}
    \theta_{j} \propto \left(\frac{L_{j}}{\Bar{\rho}R^{2}}\right)^{1/2}  
  \end{equation}  \\
 
 %considering the pressure differences in the jet and cocoon  in the context of different progenitor properties.  The pressure in the cocoon can be estimated by
 % \begin{equation}
 %      p_{a} = \rho_{a}c^{2}
 %  \end{equation}
  %\cite{Brom11} (CHECK THIS - THIS IS NOT NECESSARILY WHAT WE WANT TO USE): 
  %\begin{equation}
  %     p_{c} \approx \left(\frac{L_{j}\rho_{a}}{3\pi c}\right) % \end{equation}
  % \noindent Or, just use that the cocoon pressure is
  %\begin{equation}
  %     p_{c} = \rho_{a}(\beta_{c} c)^{2}
  % \end{equation} 
  % \noindent where $\beta_{c}$ is the velocity of the cocoon.
   
  %\noindent This can be compared to the energy density in the jet:
  % \begin{equation}
  %%     p_{j} = \frac{L_{j}}{4\Sigma_{j}c\Gamma_{j}^{2}}
  %\epsilon_{j} = \frac{L_{j}}{\Sigma_{j}c}
  % \end{equation}
 %  \noindent where $\Sigma_{j}$ is the jet cross section and $\Gamma_{j}$ is the jet Lorentz factor.  

  \begin{figure*}
  \includegraphics[width=6.5in]{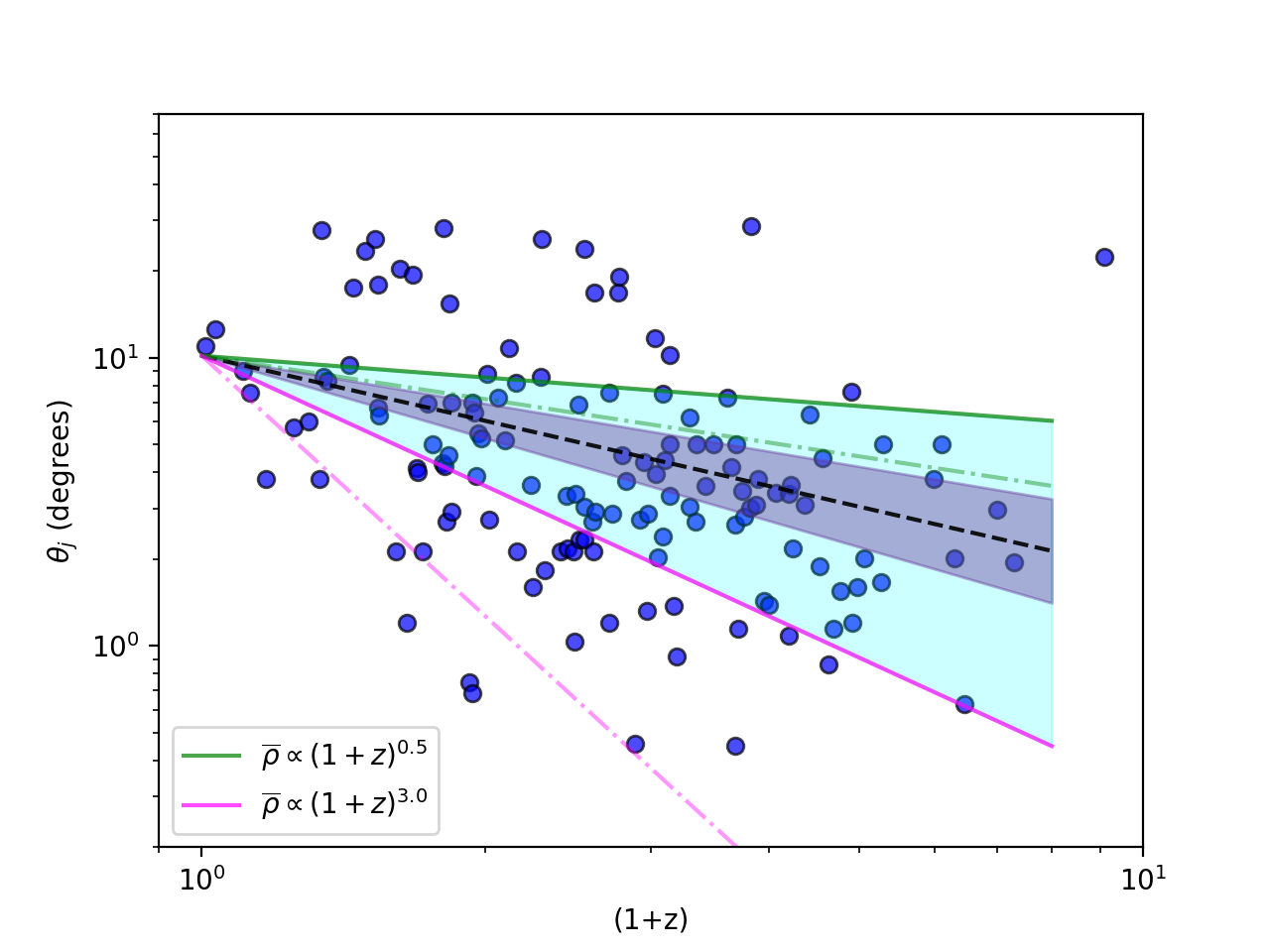}
  %{thetajz2.png} 
  \caption{Jet opening angle vs redshift. The dashed black line and purple region show the best fit to the data, $\theta_{j} \propto (1+z)^{-0.75 \pm 0.25}$.  The cyan region shows the expected relationship between $\theta_{j}$ and $(1+z)$ assuming that collimation of the jet is related to the average density of the star as $\theta_{j} \propto (1/\Bar{\rho})^{1/2}$ and that the average stellar density evolves with redshift, bounded by $(1+z)^{-0.5}$ (upper green line) and $(1+z)^{3.0}$ (lower magenta line).  The dash-dot lines show the expected redshift dependence for $\theta_{j} \propto (1/\Bar{\rho})$. }
  \label{fig:thetaz}
  \end{figure*}   
  
   \begin{figure}
  \includegraphics[width=3.5in]{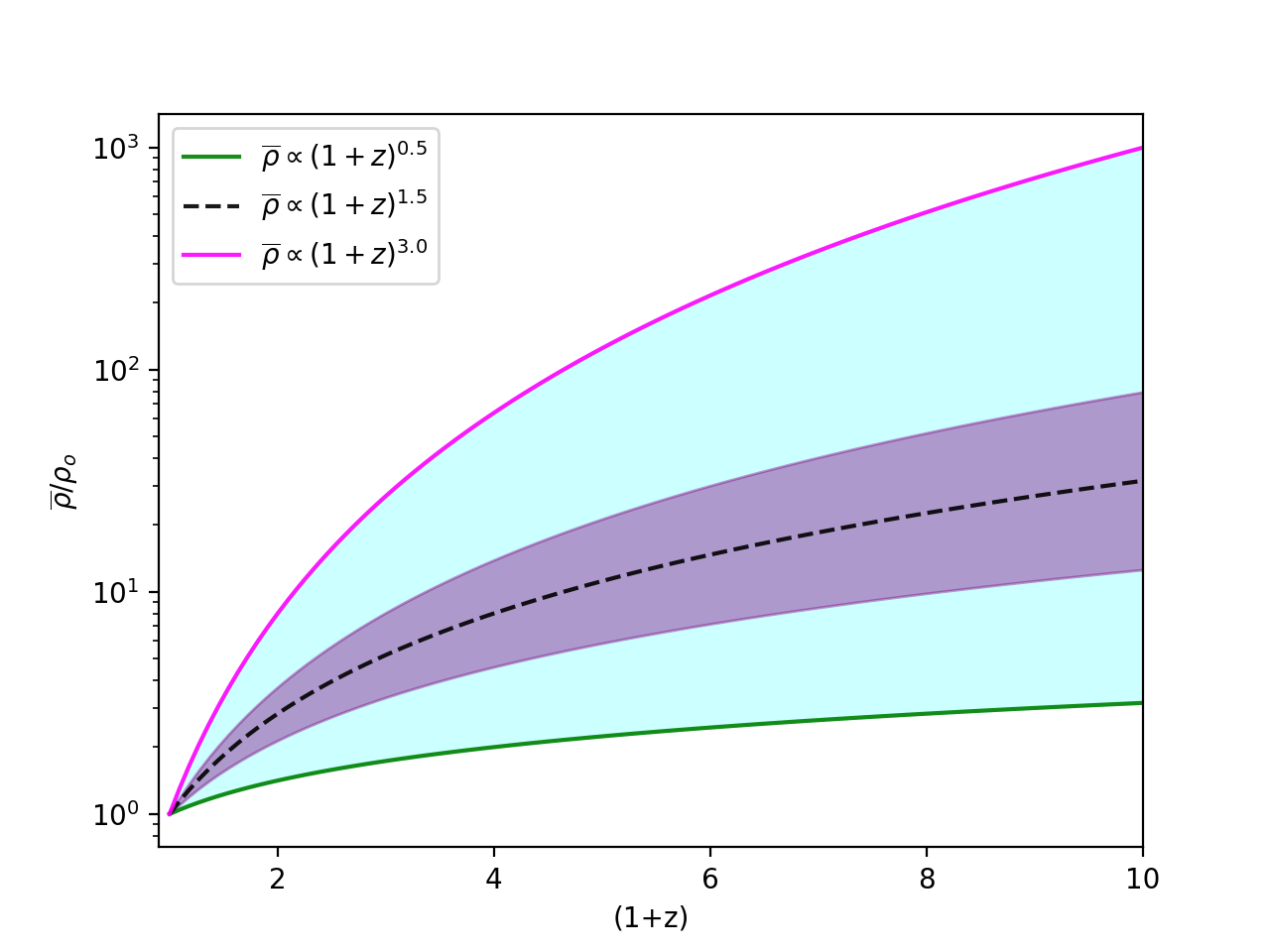} 
  \caption{Models of average stellar density as a function of redshift that may contribute the the anti-correlation between jet opening angle and redshift.  The dark shaded region shows the required density evolution to reproduce the best fit line to the data $(\theta_{j} \propto (1+z)^{-0.75 \pm 0.25})$, assuming $\theta_{j} \propto (1/\overline{\rho})^{1/2}$.}
  \label{fig:densz}
  \end{figure}     
  %%%%%% z/R
 %  Inverting equation 28 of \cite{Brom11} to solve for $(z/R)$, the scale height of the jet, we find for a jet to be collimated:
 %  \begin{equation}
 %      \frac{z_{h}}{R} < \left(\frac{L_{j}}{10^{54}\rm{erg %s^{-1}}}\right)^{\frac{1}{(2-\alpha)}}\left(\frac{R}{R_{\odot}}\right)^{\frac{1}{(2-\alpha)}}\left(\frac{\theta_{o}}{10^{\circ}}\right)^{\frac{-2}{3(2-\alpha)}}\left(\frac{10M_{\odot}}{M}\right)^{\frac{1}{(2-\alpha)}}
 %  \end{equation}
 %  \noindent This equation is too easily satisfied which is saying jets are collimated inside their stars.  This is not what I want.

%%%% Progenitor Radius   
%  What does this tell me about the progenitor radius?  It has to satisfy the following
%  \begin{equation}
 %   \frac{R}{R_{\odot}} \lesssim \frac{10^{54}\rm{erg s^{-1}}}{L_{j}} %\left(\frac{z_{h}}{R}\right)^{2-\alpha} \left(\frac{\theta_{o}}{10^{\circ}}\right)^{2/3} %\left(\frac{M}{10M_{\odot}}\right)
%  \end{equation}
   
%\subsubsection{Degree of Collimation by Stellar Envelope}   

   We can also use the formalism of \cite{MLB07} to estimate the degree of jet collimation as it travels through a stellar envelope.  Considering the ram pressure due to the deflection of a jet by the boundary layer, and under the condition of a narrow, relativistic jet, they arrive at an expression for the jet opening angle as it propagates through the star:
   \begin{equation}
       \theta_{j} = \frac{2\theta_{o}}{2 + K\theta_{o}(h^{2} - h_{o}^{2})}
   \end{equation}
   where $K = c\pi p_{c}/L_{j}$, $h$ is the height or distance from the jet launch radius and $h_{o}$ is the initial height of jet launch (note they use the variables $z$ and $z_{o}$), and $\theta_{o}$ is the initial jet opening angle (at height $h_{o}$). This equation indicates, as expected, that the farther the jet has to travel through the star, the more it is collimated.  If we consider $h \sim R >> h_{o}$ and $K\theta_{o}R^{2} >> 2$, then we find:
   \begin{equation}
    \theta_{j} \propto \left(\frac{L_{j}}{\Bar{\rho}R^{2}}\right)  
  \end{equation}  \\
  This gives a stronger dependence on average density compared to our simple comparison of energy densities (equation 7).  Nonetheless, as with equation 7, it points to the direct relationship between jet luminosity and opening angle (as discussed in \S 3) and the inverse relationship between average ambient density and jet opening angle - i.e. that denser stars collimate jets more. In what follows, we focus on $\theta_{j} \propto (1/\Bar{\rho})^{1/2}$, but we stress the qualitative behavior and results remain the same whether we use equation 7 or the stronger dependence in equation 9.
   %This is a very broad statement, however.  Looking in more detail, for example, at the angular energy distribution in the jet, \cite{MLB07} found that there is no obvious trend between this quantity (jet angular energy distribution) and progenitor model, for the massive star progenitors they explored.\\
   \\

   \subsection{Cosmological Evolution of Jet Opening Angle}
%    As discussed above and explored in detail in \cite{Matz03} and \cite{Brom11}, there are a number of factors that play a role in the collimation of the jet by the stellar envelope. The condition for collimation given in equation 8 in particular relates the luminosity of the jet to the necessary stellar parameters/density profile required to collimate this jet.
 Given equation 7 (or 9) above, we can ask whether the jet opening angle  dependence on $L_{j}$, $\Bar{\rho}$, and/or $R$ might lead to cosmological evolution of this quantity.  We have already shown \citep{LR19} that the gamma-ray emitted luminosity $L_{\gamma} (\propto L_{j}$) is not correlated with redshift.  Average density, however, is expected to evolve as both the IMF and metallicity evolve throughout cosmic time. We can estimate how the average density, $\Bar{\rho} \sim M/R^{3}$, changes with progenitor properties (e.g. mass, radius, metallicity), and how these quantities are interrelated. Massive main sequence stars follow a mass-radius relation $R \propto M^{\sim 0.6 - 0.8}$ \citep{DK91}, although a  number of things can complicate this question, like the presence of convection, magnetic fields, etc.. Therefore numerical models of massive stars can be used to estimate the average density for various progenitors.  \cite{MA09} have done this and show - using the models of \cite{WH06} - that the average stellar density for massive stars scales roughly as the stellar mass. 
   
   In that case (i.e. assuming average density scales roughly as mass and depends only weakly on radius) we find
  \begin{equation}
      \theta_{j} \propto [L_{j}/M]^{1/2}
  \end{equation}

  This suggests that {\em progenitor mass evolution could be the underlying cause of lGRB beaming angle evolution}.  Consider the effects of an evolving IMF as suggested in \cite{vD07,Dave08, WD11, Mark12} and others, and recently \cite{Leja19}, who discuss evidence for a top-heavier IMF at higher redshifts.  If average stellar density is indeed roughly proportional to stellar mass for high mass stars, then we expect higher mass stars to collimate the jet more.  If, in addition, this density evolves with redshift according to the evolution of the characteristic mass of the IMF, then stars at higher redshift will more strongly collimate GRB jets, leading to the observed anti-correlation between jet opening angle and redshift. \\

   \cite{Dave08} proposed an evolving initial stellar mass function, where the characteristic or break mass evolves according to $\tilde{m} \sim 0.5(1+z)^{2}M_{\odot}$ (see also \cite{Dave11}).
  Using equation 7 above, this would lead to a jet opening angle dependence $\theta_{j} \propto (1+z)^{-1}$, consistent with the observed correlation.  Figure~\ref{fig:thetaz} shows the jet angle-redshift relationship resulting from density (or mass) evolution ranging from  $\Bar{\rho} \propto (1+z)^{0.5}$ up to $\Bar{\rho} \propto (1+z)^{3}$ (shown in Figure~\ref{fig:densz}).    
  
  %However, again we point out that the suggested mass evolution of the IMF \citep{Dave08} leads us to an average stellar density evolution of $\Bar{\rho} \propto (1+z)^{\sim 2}$, which leads us to an average beaming angle evolution of $\theta_{j} \propto (1+z)^{-1}$, consistent with the correlation we see in the data. Again, it's hard to nail this exact relationship down for a number of complicated reasons but guidance from \citep{vD07,WD11,Leja19} helps.\\

  This of course ignores various underlying relationships between lGRB properties (and their dependence on progenitor type).  For example, if higher mass stars tend to produce more luminous jets, this will mitigate the effect of the higher average stellar density collimating the jet more (although we stress again we found no correlation between $L_{\gamma}$ and redshift).  In addition the relationship between progenitor type and the underlying jet launch - intimately connected to the angular momentum and magnetic flux of the system - is also not considered. As discussed in \S 3, these issues really require high fidelity simulations of the collapse and launch of a jet under a wide range of conditions; we explore this in a future publication (Hurtado et al., in prep).
  
  %We've also ignored the degree of sideways expansion which will occur to some extent when the jet breaks out of the star until it accelerates to a Lorentz factor $\Gamma \theta_{j} \sim 20$ where it then propagates more or less ballistically \citep{Tch10}, although - as discussed in the previous section - this may not be a large effect. {\em Note we find no correlation between the jet opening angle and estimated bulk Lorentz factor for lGRBs for which we have this data - discuss this?}  And - again - we have neglected the scatter introduced by other potential progenitor systems.  We explore this latter point in detail in a future publication.\\
  
  Finally, we have assumed a weak dependence of average density on stellar radius in our discussion above.  Consider the previous section which pointed out that, on average, the longer a jet spends in the stellar envelope environment the more it will be collimated. Therefore, higher radii stars (provided their envelopes are dense enough) would be expected to have more collimated jets (see \S 4.1). Stars with more mass will have somewhat larger radii and so, naively, again we are led to the conclusion that higher mass (presumably low metallicity) stars at higher redshifts will lead to more collimated jets, consistent with the $\theta_{j}-(1+z)$ anti-correlation.
   And indeed, if we use empirical mass-radius relationships for massive stars, $R \propto M^{\sim 0.7}$, we find that the jet opening angle depends on stellar radius as $\theta_{j} \propto (1/R)^{\sim 0.2 - 0.4}$. Note this is similar to the relationship between $\theta_{j}$ and $R$ reported in \cite{Tch10}.
   
  %Add some OOM calculations here on $\theta_{j} \propto h^{-\beta} \propto (1+z)^{-\eta}$.  We can use the results of \cite{Tch10} (equation 10) but those are sort of by construction.  We can use equation 9 taken from \cite{MLB07}, or we can also just look at the pressure balance at face value - $\theta_{j} \propto h^{-1}$.  If we can get a handle on average stellar radius as a function of redshift (account for metallicity - i.e. high redshift stars also have lower metallicity and therefore tend to be more compact), we can estimate a function.
 %Handwavy stuff I've said. Want to argue that on average the effects of increased mass (density) and/or increased radius at high redshift lead to more collimated jets at higher redshifts. Need a better mass-radius relationship for high mass, lower metallicity stars. }

%\section{ Offset Evolution}

%\section{On Short and Intermediate (Long to Short) GRBs}

%\subsection{On the Potential Prompt Duration Evolution}

 %However, we do find a mild anti correlation between these quantities which is suggestive.  Actually it's very mild and very weak, so let's not go here.  Extreme scatter in the plot.

\section{Conclusions}
 We have investigated the nature of the observed anti-correlation between long gamma-ray burst jet opening angle and redshift.  We suggest that high redshift stars producing lGRBs have higher average density and are able to collimate a GRB jet more effectively.  Higher density stars at high redshift are expected in an evolving IMF scenario, and the $\theta_{j}-(1+z)$ correlation is consistent with the quantitative form of some proposed IMF characteristic mass evolution, under the assumption that average stellar density is proportional to mass for lGRB progenitors. Therefore, {\em we assert that beaming angle evolution is a result of progenitor mass evolution} in a collapsar scenario for lGRBs. Additionally, these higher mass stars will be slightly larger in radii (for a given metallicity) and therefore the jet will spend more time in the stellar envelope being collimated (thus emerging with a narrower opening angle). In summary:
 \begin{itemize}
    \item We examine the luminosity conditions for collimation and succesful jet launch in a massive star progenitor and find the former is easily met, while the latter is met if we assume the gamma-ray radiative efficiency is low for some lGRBs (or, alternatively, that the stellar pressure is lower than what we've assumed). We find a strong correlation between gamma-ray luminosity and jet opening angle in the data, with $L_{j} \propto \theta_{j}^{1.5}$.  We suggest this is a natural selection effect - only the most luminous GRBs are able to launch jets with wide opening angles.
     \item Jet opening angle at breakout is inversely proportional to the average stellar density, $\theta_{j} \propto (1/\Bar{\rho})^{\beta}$, with $\beta$ between $1/2$ and $1$. We suggest the anti-correlation between jet opening angle and redshift can be explained if the average stellar density evolves with redshift as $\bar{\rho} \propto (1+z)^{\sim 2 \pm 1}$.  This is in line with the proposed evolution of the characteristic mass of the IMF \citep{Dave08}, if we can assume that density is roughly proportional to mass (i.e. has a weak dependence on radius), as has been shown for high mass stars \cite{MA09}.  In this case, and taking $\beta = 1/2$, we can write the jet opening angle as $\theta_{j} \propto (L_{j}/M)^{1/2}$.
   %  \item The anti-correlation may also be explained if stars at higher redshift retain more angular momentum (due to lower metallicity).
     
 \end{itemize}
 
 We have focused on exploring the beaming angle evolution in the context of a jet propagating through a massive star envelope 
 %(see also \cite{Nag12} who consider jet breakout conditions for Pop III and Wolf-Rayet lGRB progenitors).
 We have interpreted our results in the context of average stellar density as it relates to stellar mass and radius.  As discussed in \S 2.2 and throughout the paper, metallicity plays an important role in its connection to the mass and stellar structure of the lGRB progenitor.  Besides the trend of a larger population of higher-mass, lower-metallicity stars at high redshift, \cite{MA09} suggest that the cocoon is more narrowly collimated in lower metallicity stars, because of the steep drop off in density outside the surface (see their Figure 1) compared to higher metallicity stars (their Figure~\ref{fig:thetaz}).  This steep drop off allows for low-metallicity jets to be more ballistic and would therefore also lead to a narrower measured jet opening angle.   We again caution about the various selection effects that can complicate the determination of the lGRB jet opening angle; however, this paper provides a possible physical explanation, and to some extent a prediction of its existence in the case when the IMF is weighted toward higher mass, lower metallicity stars in the early universe.

A very important implication of the cosmic evolution of lGRB jet opening angle is that this relationship significantly affects the rates of lGRBs as a function of redshift, with potentially many more GRBs at high redshifts than current estimates indicate.  Even without accounting for beaming angle evolution, several studies (e.g. \cite{LRFRR02, Kist08, Yuk08, Kist09}) suggest that the high redshift star formation rate - under certain assumptions about its connection to the lGRB population - is higher than rates estimated from other stellar populations. Our results amplify this effect. That is, a smaller opening angle at higher redshift implies that a smaller fraction of lGRBs are observed.  Thus, the lGRB rate at high redshift may be substantially underestimated in the standard models that assume a single, non-evolving jet opening angle - possibly up to an order of magnitude underestimate of the GRB rate at the highest redshifts (see  Figure 7 of \cite{Hop06} who show rate densities of core-collapse supernovae out to high redshifts for different IMFs).  This also supports the suggestion that the IMF is more top-heavy at higher redshifts, as it implies a larger fraction of the total stellar mass is in massive GRB progenitor stars at higher redshift. We examine this issue in detail in an upcoming publication; future missions such as THESEUS \citep{Am2019}, with potential to detect GRBs out to a redshift of 12, may help us test these suggestions and perhaps more definitively and ultimately determine the relationship between the GRB rate and star formation rate.  \\

 As discussed in \S 2, the anti-correlation between jet opening angle and redshift may also be explained a number of other ways for a given progenitor, including through a dependence on the amount of angular momentum retained in the star as it forms a GRB, the magnetic field strength and geometry, and/or other details of the jet launching mechanism.  Additionally, we might account for the observed beaming angle evolution if different progenitor systems are be dominating the population at different redshifts - for example, \cite{MA09} find that collapsar jets are more narrowly beamed than jets from mergers which might indicate that collapsars are dominating the progenitor distribution at high redshift while binary scenarios dominate at low redshift (although the presence of supernovae in the light curves of long GRBs will constrain potential models and will necessitate association with a massive star in some way).  The different rates of possible lGRB progenitors (e.g. Table 1 of \cite{Lev16}) and their distinct signatures as a function of redshift are important questions that should be understood if we are to understand the observations of the global lGRB population \citep{Kist08}.  We speculate that the scatter we see in the $\theta_{j}-(1+z)$ correlation could be a consequence of this (different progenitors) in addition to variation among properties of single progenitor type.  We explore this further in an upcoming publication (Aykutalp et al., in prep).\\

 Finally, we mention that \cite{LR19}  also suggested there exists an anti-correlation between the intrinsic prompt duration and redshift.  This trend was also seen in a radio bright sub-sample of energetic long GRBs, but {\em not} in a radio dark sample \citep{LR19c}, hinting at the possibility of a progenitor signature/effect.   Intrinsic duration of the prompt emission is a reflection of the lifetime of the disk, which in turn reflects the mass and accretion rate of the disk (although a number of external effects can affect this duration by up to a factor of two or so, e.g. \cite{GM15}).  Again, all of these things relate to fundamental properties of the progenitor system, and should be understood in the context of lGRB progenitor models.  As observations and simulations of these systems improve, we can test our understanding of these trends, with the hope of learning something truly fundamental about the lGRB progenitor.
  
  \section{Acknowledgements}
   We are very grateful to the anonymous referee for valuable and helpful comments on the manuscript. We also thank Mohira Rassel, Sanjana Curtis, and Brooke Polak for enlightening discussions.
  This work was supported by the US Department of Energy through the Los Alamos National Laboratory.  Los Alamos National Laboratory is operated by Triad National Security, LLC, for the National Nuclear Security Administration of U.S. Department of Energy (Contract No. 89233218CNA000001). J. ~L. ~J. and A.~A. are  supported  by  a  LANL  LDRD Exploratory  Research  Grant  20170317ER.  LA-UR-19-31599

\bibliographystyle{mnras}
\bibliography{refs} % if your bibtex file is called example.bib

\begin{thebibliography}{}
\makeatletter
\relax
\def\mn@urlcharsother{\let\do\@makeother \do\$\do\&\do\#\do\^\do\_\do\%\do\~}
\def\mn@doi{\begingroup\mn@urlcharsother \@ifnextchar [ {\mn@doi@}
  {\mn@doi@[]}}
\def\mn@doi@[#1]#2{\def\@tempa{#1}\ifx\@tempa\@empty \href
  {http://dx.doi.org/#2} {doi:#2}\else \href {http://dx.doi.org/#2} {#1}\fi
  \endgroup}
\def\mn@eprint#1#2{\mn@eprint@#1:#2::\@nil}
\def\mn@eprint@arXiv#1{\href {http://arxiv.org/abs/#1} {{\tt arXiv:#1}}}
\def\mn@eprint@dblp#1{\href {http://dblp.uni-trier.de/rec/bibtex/#1.xml}
  {dblp:#1}}
\def\mn@eprint@#1:#2:#3:#4\@nil{\def\@tempa {#1}\def\@tempb {#2}\def\@tempc
  {#3}\ifx \@tempc \@empty \let \@tempc \@tempb \let \@tempb \@tempa \fi \ifx
  \@tempb \@empty \def\@tempb {arXiv}\fi \@ifundefined
  {mn@eprint@\@tempb}{\@tempb:\@tempc}{\expandafter \expandafter \csname
  mn@eprint@\@tempb\endcsname \expandafter{\@tempc}}}

\bibitem[\protect\citeauthoryear{{Abbott} et~al.,}{{Abbott}
  et~al.}{2017}]{Ab17}
{Abbott} B.~P.,  et~al., 2017, \mn@doi [Physical Review Letters]
  {10.1103/PhysRevLett.119.161101}, \href
  {http://adsabs.harvard.edu/abs/2017PhRvL.119p1101A} {119, 161101}

\bibitem[\protect\citeauthoryear{{Aloy}, {Cuesta-Mart{\'{\i}}nez}  \&
  {Obergaulinger}}{{Aloy} et~al.}{2018}]{Aloy18}
{Aloy} M.~A.,  {Cuesta-Mart{\'{\i}}nez} C.,   {Obergaulinger} M.,  2018,
  \mn@doi [\mnras] {10.1093/mnras/sty1212}, \href
  {https://ui.adsabs.harvard.edu/abs/2018MNRAS.478.3576A} {478, 3576}

\bibitem[\protect\citeauthoryear{{Amati}, {O'Brien}, {Gotz}  \&
  {Bozzo}}{{Amati} et~al.}{2019}]{Am2019}
{Amati} L.,  {O'Brien} P.,  {Gotz} D.,   {Bozzo} E.,  2019, in AAS/High Energy
  Astrophysics Division. AAS/High Energy Astrophysics Division.
p. 303.02

\bibitem[\protect\citeauthoryear{{Barkov} \& {Komissarov}}{{Barkov} \&
  {Komissarov}}{2008}]{BK08}
{Barkov} M.~V.,  {Komissarov} S.~S.,  2008, \mn@doi [International Journal of
  Modern Physics D] {10.1142/S0218271808013285}, \href
  {http://adsabs.harvard.edu/abs/2008IJMPD..17.1669B} {17, 1669}

\bibitem[\protect\citeauthoryear{{Barkov} \& {Komissarov}}{{Barkov} \&
  {Komissarov}}{2010}]{BK10}
{Barkov} M.~V.,  {Komissarov} S.~S.,  2010, \mn@doi [\mnras]
  {10.1111/j.1365-2966.2009.15792.x}, \href
  {https://ui.adsabs.harvard.edu/abs/2010MNRAS.401.1644B} {401, 1644}

\bibitem[\protect\citeauthoryear{{Belczynski}, {Bulik}  \&
  {Rudak}}{{Belczynski} et~al.}{2002}]{BBR02}
{Belczynski} K.,  {Bulik} T.,   {Rudak} B.,  2002, \mn@doi [\apj]
  {10.1086/339860}, \href
  {https://ui.adsabs.harvard.edu/abs/2002ApJ...571..394B} {571, 394}

\bibitem[\protect\citeauthoryear{{Berger}}{{Berger}}{2014}]{Berg14}
{Berger} E.,  2014, \mn@doi [\araa] {10.1146/annurev-astro-081913-035926},
  \href {http://adsabs.harvard.edu/abs/2014ARA%26A..52...43B} {52, 43}

\bibitem[\protect\citeauthoryear{{Blandford} \& {Znajek}}{{Blandford} \&
  {Znajek}}{1977}]{BZ77}
{Blandford} R.~D.,  {Znajek} R.~L.,  1977, \mn@doi [\mnras]
  {10.1093/mnras/179.3.433}, \href
  {https://ui.adsabs.harvard.edu/abs/1977MNRAS.179..433B} {179, 433}

\bibitem[\protect\citeauthoryear{{Bloom}, {Kulkarni}  \& {Djorgovski}}{{Bloom}
  et~al.}{2002}]{BKD02}
{Bloom} J.~S.,  {Kulkarni} S.~R.,   {Djorgovski} S.~G.,  2002, \mn@doi [\aj]
  {10.1086/338893}, \href {http://adsabs.harvard.edu/abs/2002AJ....123.1111B}
  {123, 1111}

\bibitem[\protect\citeauthoryear{{Bromberg}, {Nakar}, {Piran}  \&
  {Sari}}{{Bromberg} et~al.}{2011}]{Brom11}
{Bromberg} O.,  {Nakar} E.,  {Piran} T.,   {Sari} R.,  2011, \mn@doi [\apj]
  {10.1088/0004-637X/740/2/100}, \href
  {https://ui.adsabs.harvard.edu/abs/2011ApJ...740..100B} {740, 100}

\bibitem[\protect\citeauthoryear{{Bromm} \& {Loeb}}{{Bromm} \&
  {Loeb}}{2006}]{BL06}
{Bromm} V.,  {Loeb} A.,  2006, \mn@doi [\apj] {10.1086/500799}, \href
  {https://ui.adsabs.harvard.edu/abs/2006ApJ...642..382B} {642, 382}

\bibitem[\protect\citeauthoryear{{Burrows}, {Dessart}, {Livne}, {Ott}  \&
  {Murphy}}{{Burrows} et~al.}{2007}]{Burr07}
{Burrows} A.,  {Dessart} L.,  {Livne} E.,  {Ott} C.~D.,   {Murphy} J.,  2007,
  \mn@doi [\apj] {10.1086/519161}, \href
  {https://ui.adsabs.harvard.edu/abs/2007ApJ...664..416B} {664, 416}

\bibitem[\protect\citeauthoryear{{Callingham}, {Tuthill}, {Pope}, {Williams},
  {Crowther}, {Edwards}, {Norris}  \& {Kedziora-Chudczer}}{{Callingham}
  et~al.}{2019}]{Call19}
{Callingham} J.~R.,  {Tuthill} P.~G.,  {Pope} B.~J.~S.,  {Williams} P.~M.,
  {Crowther} P.~A.,  {Edwards} M.,  {Norris} B.,   {Kedziora-Chudczer} L.,
  2019, \mn@doi [Nature Astronomy] {10.1038/s41550-018-0617-7}, \href
  {https://ui.adsabs.harvard.edu/abs/2019NatAs...3...82C} {3, 82}

\bibitem[\protect\citeauthoryear{{Cantiello}, {Yoon}, {Langer}  \&
  {Livio}}{{Cantiello} et~al.}{2007}]{Cant07}
{Cantiello} M.,  {Yoon} S.~C.,  {Langer} N.,   {Livio} M.,  2007, \mn@doi
  [\aap] {10.1051/0004-6361:20077115}, \href
  {https://ui.adsabs.harvard.edu/abs/2007A&A...465L..29C} {465, L29}

\bibitem[\protect\citeauthoryear{{Chrimes}, {Stanway}  \& {Eldridge}}{{Chrimes}
  et~al.}{2019}]{CSE19}
{Chrimes} A.~A.,  {Stanway} E.~R.,   {Eldridge} J.~J.,  2019, arXiv e-prints,
  \href {https://ui.adsabs.harvard.edu/abs/2019arXiv191108387C} {p.
  arXiv:1911.08387}

\bibitem[\protect\citeauthoryear{{Chruslinska}, {Jerabkova}, {Nelemans}  \&
  {Yan}}{{Chruslinska} et~al.}{2020}]{Chrus20}
{Chruslinska} M.,  {Jerabkova} T.,  {Nelemans} G.,   {Yan} Z.,  2020, arXiv
  e-prints, \href {https://ui.adsabs.harvard.edu/abs/2020arXiv200211122C} {p.
  arXiv:2002.11122}

\bibitem[\protect\citeauthoryear{{D'Avanzo}}{{D'Avanzo}}{2015}]{DAvanz15}
{D'Avanzo} P.,  2015, \mn@doi [Journal of High Energy Astrophysics]
  {10.1016/j.jheap.2015.07.002}, \href
  {http://adsabs.harvard.edu/abs/2015JHEAp...7...73D} {7, 73}

\bibitem[\protect\citeauthoryear{{Dainotti}, {Del Vecchio}  \&
  {Tarnopolski}}{{Dainotti} et~al.}{2018}]{Dai18}
{Dainotti} M.~G.,  {Del Vecchio} R.,   {Tarnopolski} M.,  2018, \mn@doi
  [Advances in Astronomy] {10.1155/2018/4969503}, \href
  {https://ui.adsabs.harvard.edu/abs/2018AdAst2018E...1D} {2018, 4969503}

\bibitem[\protect\citeauthoryear{{Dav{\'e}}}{{Dav{\'e}}}{2008}]{Dave08}
{Dav{\'e}} R.,  2008, \mn@doi [\mnras] {10.1111/j.1365-2966.2008.12866.x},
  \href {https://ui.adsabs.harvard.edu/abs/2008MNRAS.385..147D} {385, 147}

\bibitem[\protect\citeauthoryear{{Dav{\'e}}}{{Dav{\'e}}}{2011}]{Dave11}
{Dav{\'e}} R.,  2011, in {Treyer} M.,  {Wyder} T.,  {Neill} J.,  {Seibert} M.,
   {Lee} J.,  eds,  Astronomical Society of the Pacific Conference Series Vol.
  440, UP2010: Have Observations Revealed a Variable Upper End of the Initial
  Mass Function?. p.~353 (\mn@eprint {arXiv} {1008.5283})

\bibitem[\protect\citeauthoryear{{Davies}, {Levan}, {Larsson}, {King}  \&
  {Fruchter}}{{Davies} et~al.}{2007}]{Dav07}
{Davies} M.~B.,  {Levan} A.~J.,  {Larsson} J.,  {King} A.~R.,   {Fruchter}
  A.~S.,  2007, in {Axelsson} M.,  {Ryde} F.,  eds,  American Institute of
  Physics Conference Series Vol. 906, Gamma-Ray Bursts: Prospects for GLAST. pp
  69--78 (\mn@eprint {arXiv} {0704.1899}), \mn@doi{10.1063/1.2737408}

\bibitem[\protect\citeauthoryear{{Della Valle} et~al.,}{{Della Valle}
  et~al.}{2006}]{DV06}
{Della Valle} M.,  et~al., 2006, \mn@doi [\nat] {10.1038/nature05374}, \href
  {https://ui.adsabs.harvard.edu/abs/2006Natur.444.1050D} {444, 1050}

\bibitem[\protect\citeauthoryear{{Demircan} \& {Kahraman}}{{Demircan} \&
  {Kahraman}}{1991}]{DK91}
{Demircan} O.,  {Kahraman} G.,  1991, \mn@doi [\apss] {10.1007/BF00639097},
  \href {https://ui.adsabs.harvard.edu/abs/1991Ap&SS.181..313D} {181, 313}

\bibitem[\protect\citeauthoryear{{Deng}, {Wang}, {Guo}, {Lu}, {Wang}, {Wei},
  {Wu}  \& {Liang}}{{Deng} et~al.}{2016}]{Deng16}
{Deng} C.-M.,  {Wang} X.-G.,  {Guo} B.-B.,  {Lu} R.-J.,  {Wang} Y.-Z.,  {Wei}
  J.-J.,  {Wu} X.-F.,   {Liang} E.-W.,  2016, \mn@doi [\apj]
  {10.3847/0004-637X/820/1/66}, \href
  {http://adsabs.harvard.edu/abs/2016ApJ...820...66D} {820, 66}

\bibitem[\protect\citeauthoryear{{Efron} \& {Petrosian}}{{Efron} \&
  {Petrosian}}{1992}]{EP92}
{Efron} B.,  {Petrosian} V.,  1992, \mn@doi [\apj] {10.1086/171931}, \href
  {http://adsabs.harvard.edu/abs/1992ApJ...399..345E} {399, 345}

\bibitem[\protect\citeauthoryear{{Fong} \& {Berger}}{{Fong} \&
  {Berger}}{2013}]{Fong2013}
{Fong} W.,  {Berger} E.,  2013, \mn@doi [\apj] {10.1088/0004-637X/776/1/18},
  \href {http://adsabs.harvard.edu/abs/2013ApJ...776...18F} {776, 18}

\bibitem[\protect\citeauthoryear{{Fong}, {Berger}  \& {Fox}}{{Fong}
  et~al.}{2010}]{Fong2010}
{Fong} W.,  {Berger} E.,   {Fox} D.~B.,  2010, \mn@doi [\apj]
  {10.1088/0004-637X/708/1/9}, \href
  {http://adsabs.harvard.edu/abs/2010ApJ...708....9F} {708, 9}

\bibitem[\protect\citeauthoryear{{Fryer} \& {Heger}}{{Fryer} \&
  {Heger}}{2005}]{FH05}
{Fryer} C.~L.,  {Heger} A.,  2005, \mn@doi [\apj] {10.1086/428379}, \href
  {https://ui.adsabs.harvard.edu/abs/2005ApJ...623..302F} {623, 302}

\bibitem[\protect\citeauthoryear{{Fryer} \& {Woosley}}{{Fryer} \&
  {Woosley}}{1998}]{FW98}
{Fryer} C.~L.,  {Woosley} S.~E.,  1998, \mn@doi [\apjl] {10.1086/311493}, \href
  {https://ui.adsabs.harvard.edu/abs/1998ApJ...502L...9F} {502, L9}

\bibitem[\protect\citeauthoryear{{Fryer}, {Belczynski}, {Berger}, {Th{\"o}ne},
  {Ellinger}  \& {Bulik}}{{Fryer} et~al.}{2013}]{Fry13}
{Fryer} C.~L.,  {Belczynski} K.,  {Berger} E.,  {Th{\"o}ne} C.,  {Ellinger} C.,
    {Bulik} T.,  2013, \mn@doi [\apj] {10.1088/0004-637X/764/2/181}, \href
  {https://ui.adsabs.harvard.edu/abs/2013ApJ...764..181F} {764, 181}

\bibitem[\protect\citeauthoryear{{Fynbo} et~al.,}{{Fynbo} et~al.}{2006}]{Fyn06}
{Fynbo} J.~P.~U.,  et~al., 2006, \mn@doi [\nat] {10.1038/nature05375}, \href
  {https://ui.adsabs.harvard.edu/abs/2006Natur.444.1047F} {444, 1047}

\bibitem[\protect\citeauthoryear{{Gal-Yam} et~al.,}{{Gal-Yam}
  et~al.}{2006}]{GY06}
{Gal-Yam} A.,  et~al., 2006, \mn@doi [\nat] {10.1038/nature05373}, \href
  {https://ui.adsabs.harvard.edu/abs/2006Natur.444.1053G} {444, 1053}

\bibitem[\protect\citeauthoryear{{Galama} et~al.,}{{Galama}
  et~al.}{1998}]{Gal98}
{Galama} T.~J.,  et~al., 1998, \mn@doi [\nat] {10.1038/27150}, \href
  {https://ui.adsabs.harvard.edu/abs/1998Natur.395..670G} {395, 670}

\bibitem[\protect\citeauthoryear{{Gao} \& {M{\'e}sz{\'a}ros}}{{Gao} \&
  {M{\'e}sz{\'a}ros}}{2015}]{GM15}
{Gao} H.,  {M{\'e}sz{\'a}ros} P.,  2015, \mn@doi [\apj]
  {10.1088/0004-637X/802/2/90}, \href
  {http://adsabs.harvard.edu/abs/2015ApJ...802...90G} {802, 90}

\bibitem[\protect\citeauthoryear{{Gehrels} et~al.,}{{Gehrels}
  et~al.}{2006}]{Gehr06}
{Gehrels} N.,  et~al., 2006, \mn@doi [\nat] {10.1038/nature05376}, \href
  {https://ui.adsabs.harvard.edu/abs/2006Natur.444.1044G} {444, 1044}

\bibitem[\protect\citeauthoryear{{Gehrels}, {Ramirez-Ruiz}  \& {Fox}}{{Gehrels}
  et~al.}{2009}]{GRRF09}
{Gehrels} N.,  {Ramirez-Ruiz} E.,   {Fox} D.~B.,  2009, \mn@doi [\araa]
  {10.1146/annurev.astro.46.060407.145147}, \href
  {http://adsabs.harvard.edu/abs/2009ARA%26A..47..567G} {47, 567}

\bibitem[\protect\citeauthoryear{{Hamidani}, {Kiuchi}  \& {Ioka}}{{Hamidani}
  et~al.}{2019}]{HKI19}
{Hamidani} H.,  {Kiuchi} K.,   {Ioka} K.,  2019, arXiv e-prints, \href
  {https://ui.adsabs.harvard.edu/abs/2019arXiv190905867H} {}

\bibitem[\protect\citeauthoryear{Hjorth \& Bloom}{Hjorth \& Bloom}{2012}]{HB12}
Hjorth J.,  Bloom J.~S.,  2012, Gamma-ray bursts

\bibitem[\protect\citeauthoryear{{Hjorth} et~al.,}{{Hjorth}
  et~al.}{2003}]{Hjorth03}
{Hjorth} J.,  et~al., 2003, \mn@doi [\nat] {10.1038/nature01750}, \href
  {http://adsabs.harvard.edu/abs/2003Natur.423..847H} {423, 847}

\bibitem[\protect\citeauthoryear{{Hopkins} \& {Beacom}}{{Hopkins} \&
  {Beacom}}{2006}]{Hop06}
{Hopkins} A.~M.,  {Beacom} J.~F.,  2006, \mn@doi [\apj] {10.1086/506610}, \href
  {https://ui.adsabs.harvard.edu/abs/2006ApJ...651..142H} {651, 142}

\bibitem[\protect\citeauthoryear{{Kelly}}{{Kelly}}{2007}]{Kell07}
{Kelly} B.~C.,  2007, \mn@doi [\apj] {10.1086/519947}, \href
  {https://ui.adsabs.harvard.edu/abs/2007ApJ...665.1489K} {665, 1489}

\bibitem[\protect\citeauthoryear{{King}, {Olsson}  \& {Davies}}{{King}
  et~al.}{2007}]{King07}
{King} A.,  {Olsson} E.,   {Davies} M.~B.,  2007, \mn@doi [\mnras]
  {10.1111/j.1745-3933.2006.00259.x}, \href
  {https://ui.adsabs.harvard.edu/abs/2007MNRAS.374L..34K} {374, L34}

\bibitem[\protect\citeauthoryear{{Kinugawa} \& {Asano}}{{Kinugawa} \&
  {Asano}}{2017}]{KA17}
{Kinugawa} T.,  {Asano} K.,  2017, \mn@doi [\apjl] {10.3847/2041-8213/aa95bb},
  \href {https://ui.adsabs.harvard.edu/abs/2017ApJ...849L..29K} {849, L29}

\bibitem[\protect\citeauthoryear{{Kistler}, {Y{\"u}ksel}, {Beacom}  \&
  {Stanek}}{{Kistler} et~al.}{2008}]{Kist08}
{Kistler} M.~D.,  {Y{\"u}ksel} H.,  {Beacom} J.~F.,   {Stanek} K.~Z.,  2008,
  \mn@doi [\apjl] {10.1086/527671}, \href
  {https://ui.adsabs.harvard.edu/abs/2008ApJ...673L.119K} {673, L119}

\bibitem[\protect\citeauthoryear{{Kistler}, {Y{\"u}ksel}, {Beacom}, {Hopkins}
  \& {Wyithe}}{{Kistler} et~al.}{2009}]{Kist09}
{Kistler} M.~D.,  {Y{\"u}ksel} H.,  {Beacom} J.~F.,  {Hopkins} A.~M.,
  {Wyithe} J. S.~B.,  2009, \mn@doi [\apjl] {10.1088/0004-637X/705/2/L104},
  \href {https://ui.adsabs.harvard.edu/abs/2009ApJ...705L.104K} {705, L104}

\bibitem[\protect\citeauthoryear{Kitagawa, Nybom  \& Stuhler}{Kitagawa
  et~al.}{2018}]{Kit18}
Kitagawa T.,  Nybom M.,   Stuhler J.,  2018, Technical report, Measurement
  error and rank correlations.
cemmap working paper

\bibitem[\protect\citeauthoryear{{Kobulnicky} et~al.,}{{Kobulnicky}
  et~al.}{2014}]{Kob14}
{Kobulnicky} H.~A.,  et~al., 2014, \mn@doi [\apjs]
  {10.1088/0067-0049/213/2/34}, \href
  {https://ui.adsabs.harvard.edu/abs/2014ApJS..213...34K} {213, 34}

\bibitem[\protect\citeauthoryear{{Kocevski} \& {Liang}}{{Kocevski} \&
  {Liang}}{2006}]{KL06}
{Kocevski} D.,  {Liang} E.,  2006, \mn@doi [\apj] {10.1086/500816}, \href
  {http://adsabs.harvard.edu/abs/2006ApJ...642..371K} {642, 371}

\bibitem[\protect\citeauthoryear{{Komissarov} \& {Barkov}}{{Komissarov} \&
  {Barkov}}{2009}]{KB09}
{Komissarov} S.~S.,  {Barkov} M.~V.,  2009, \mn@doi [\mnras]
  {10.1111/j.1365-2966.2009.14831.x}, \href
  {https://ui.adsabs.harvard.edu/abs/2009MNRAS.397.1153K} {397, 1153}

\bibitem[\protect\citeauthoryear{{Kroupa}}{{Kroupa}}{2019}]{Kroupa19}
{Kroupa} P.,  2019, arXiv e-prints, \href
  {https://ui.adsabs.harvard.edu/abs/2019arXiv191006971K} {p. arXiv:1910.06971}

\bibitem[\protect\citeauthoryear{{Kumar}, {Narayan}  \& {Johnson}}{{Kumar}
  et~al.}{2008a}]{KNJ08a}
{Kumar} P.,  {Narayan} R.,   {Johnson} J.~L.,  2008a, \mn@doi [Science]
  {10.1126/science.1159003}, \href
  {http://adsabs.harvard.edu/abs/2008Sci...321..376K} {321, 376}

\bibitem[\protect\citeauthoryear{{Kumar}, {Narayan}  \& {Johnson}}{{Kumar}
  et~al.}{2008b}]{KNJ08b}
{Kumar} P.,  {Narayan} R.,   {Johnson} J.~L.,  2008b, \mn@doi [\mnras]
  {10.1111/j.1365-2966.2008.13493.x}, \href
  {http://adsabs.harvard.edu/abs/2008MNRAS.388.1729K} {388, 1729}

\bibitem[\protect\citeauthoryear{{Laskar} et~al.,}{{Laskar}
  et~al.}{2014}]{Las14}
{Laskar} T.,  et~al., 2014, \mn@doi [\apj] {10.1088/0004-637X/781/1/1}, \href
  {https://ui.adsabs.harvard.edu/abs/2014ApJ...781....1L} {781, 1}

\bibitem[\protect\citeauthoryear{{Laskar}, {Berger}, {Chornock}, {Margutti},
  {Fong}  \& {Zauderer}}{{Laskar} et~al.}{2018a}]{Las18}
{Laskar} T.,  {Berger} E.,  {Chornock} R.,  {Margutti} R.,  {Fong} W.-f.,
  {Zauderer} B.~A.,  2018a, \mn@doi [\apj] {10.3847/1538-4357/aab8f5}, \href
  {https://ui.adsabs.harvard.edu/abs/2018ApJ...858...65L} {858, 65}

\bibitem[\protect\citeauthoryear{{Laskar} et~al.,}{{Laskar}
  et~al.}{2018b}]{Las18b}
{Laskar} T.,  et~al., 2018b, \mn@doi [\apj] {10.3847/1538-4357/aabfd8}, \href
  {https://ui.adsabs.harvard.edu/abs/2018ApJ...859..134L} {859, 134}

\bibitem[\protect\citeauthoryear{{Lee} \& {Ramirez-Ruiz}}{{Lee} \&
  {Ramirez-Ruiz}}{2007}]{LRR07}
{Lee} W.~H.,  {Ramirez-Ruiz} E.,  2007, \mn@doi [New Journal of Physics]
  {10.1088/1367-2630/9/1/017}, \href
  {http://adsabs.harvard.edu/abs/2007NJPh....9...17L} {9, 17}

\bibitem[\protect\citeauthoryear{{Leja}, {Speagle}, {Johnson}, {Conroy}, {van
  Dokkum}  \& {Franx}}{{Leja} et~al.}{2019}]{Leja19}
{Leja} J.,  {Speagle} J.~S.,  {Johnson} B.~D.,  {Conroy} C.,  {van Dokkum} P.,
   {Franx} M.,  2019, arXiv e-prints, \href
  {https://ui.adsabs.harvard.edu/abs/2019arXiv191004168L} {}

\bibitem[\protect\citeauthoryear{{Levan}, {Crowther}, {de Grijs}, {Langer},
  {Xu}  \& {Yoon}}{{Levan} et~al.}{2016}]{Lev16}
{Levan} A.,  {Crowther} P.,  {de Grijs} R.,  {Langer} N.,  {Xu} D.,   {Yoon}
  S.-C.,  2016, \mn@doi [\ssr] {10.1007/s11214-016-0312-x}, \href
  {https://ui.adsabs.harvard.edu/abs/2016SSRv..202...33L} {202, 33}

\bibitem[\protect\citeauthoryear{{Lloyd-Ronning}, {Fryer}  \&
  {Ramirez-Ruiz}}{{Lloyd-Ronning} et~al.}{2002}]{LRFRR02}
{Lloyd-Ronning} N.~M.,  {Fryer} C.~L.,   {Ramirez-Ruiz} E.,  2002, \mn@doi
  [\apj] {10.1086/341059}, \href
  {http://adsabs.harvard.edu/abs/2002ApJ...574..554L} {574, 554}

\bibitem[\protect\citeauthoryear{{Lloyd-Ronning}, {Fryer}, {Miller}, {Prasad},
  {Torres}  \& {Martin}}{{Lloyd-Ronning} et~al.}{2019a}]{LR19b}
{Lloyd-Ronning} N.~M.,  {Fryer} C.,  {Miller} J.~M.,  {Prasad} N.,  {Torres}
  C.,   {Martin} P.,  2019a, \mn@doi [\mnras] {10.1093/mnras/stz390}, \href
  {https://ui.adsabs.harvard.edu/abs/2019MNRAS.485..203L} {485, 203}

\bibitem[\protect\citeauthoryear{{Lloyd-Ronning}, {Aykutalp}  \&
  {Johnson}}{{Lloyd-Ronning} et~al.}{2019b}]{LR19}
{Lloyd-Ronning} N.~M.,  {Aykutalp} A.,   {Johnson} J.~L.,  2019b, \mn@doi
  [\mnras] {10.1093/mnras/stz2155}, \href
  {https://ui.adsabs.harvard.edu/abs/2019MNRAS.488.5823L} {488, 5823}

\bibitem[\protect\citeauthoryear{{Lloyd-Ronning}, {Gompertz}, {Pe'er},
  {Dainotti}  \& {Fruchter}}{{Lloyd-Ronning} et~al.}{2019c}]{LR19c}
{Lloyd-Ronning} N.~M.,  {Gompertz} B.,  {Pe'er} A.,  {Dainotti} M.,
  {Fruchter} A.,  2019c, \mn@doi [\apj] {10.3847/1538-4357/aaf6ac}, \href
  {https://ui.adsabs.harvard.edu/abs/2019ApJ...871..118L} {871, 118}

\bibitem[\protect\citeauthoryear{{Lloyd}, {Petrosian}  \& {Mallozzi}}{{Lloyd}
  et~al.}{2000}]{LPM00}
{Lloyd} N.~M.,  {Petrosian} V.,   {Mallozzi} R.~S.,  2000, \mn@doi [\apj]
  {10.1086/308742}, \href {http://adsabs.harvard.edu/abs/2000ApJ...534..227L}
  {534, 227}

\bibitem[\protect\citeauthoryear{L{\"u}, Zou, Lei, Zhang, Wu, Wang, Liang  \&
  L{\"u}}{L{\"u} et~al.}{2012}]{Lu12}
L{\"u} J.,  Zou Y.-C.,  Lei W.-H.,  Zhang B.,  Wu Q.,  Wang D.-X.,  Liang
  E.-W.,   L{\"u} H.-J.,  2012, The Astrophysical Journal, 751, 49

\bibitem[\protect\citeauthoryear{{Lyman} et~al.,}{{Lyman} et~al.}{2017}]{Ly17}
{Lyman} J.~D.,  et~al., 2017, \mn@doi [\mnras] {10.1093/mnras/stx220}, \href
  {http://adsabs.harvard.edu/abs/2017MNRAS.467.1795L} {467, 1795}

\bibitem[\protect\citeauthoryear{{Lynden-Bell}}{{Lynden-Bell}}{1971}]{LB71}
{Lynden-Bell} D.,  1971, \mn@doi [\mnras] {10.1093/mnras/155.1.95}, \href
  {http://adsabs.harvard.edu/abs/1971MNRAS.155...95L} {155, 95}

\bibitem[\protect\citeauthoryear{{MacFadyen} \& {Woosley}}{{MacFadyen} \&
  {Woosley}}{1999}]{MW99}
{MacFadyen} A.~I.,  {Woosley} S.~E.,  1999, \mn@doi [\apj] {10.1086/307790},
  \href {https://ui.adsabs.harvard.edu/abs/1999ApJ...524..262M} {524, 262}

\bibitem[\protect\citeauthoryear{{Mandel}, {Farr}  \& {Gair}}{{Mandel}
  et~al.}{2019}]{MFG19}
{Mandel} I.,  {Farr} W.~M.,   {Gair} J.~R.,  2019, \mn@doi [\mnras]
  {10.1093/mnras/stz896}, \href
  {https://ui.adsabs.harvard.edu/abs/2019MNRAS.486.1086M} {486, 1086}

\bibitem[\protect\citeauthoryear{{Marks}, {Kroupa}, {Dabringhausen}  \&
  {Pawlowski}}{{Marks} et~al.}{2012}]{Mark12}
{Marks} M.,  {Kroupa} P.,  {Dabringhausen} J.,   {Pawlowski} M.~S.,  2012,
  \mn@doi [\mnras] {10.1111/j.1365-2966.2012.20767.x}, \href
  {https://ui.adsabs.harvard.edu/abs/2012MNRAS.422.2246M} {422, 2246}

\bibitem[\protect\citeauthoryear{{Matzner}}{{Matzner}}{2003}]{Matz03}
{Matzner} C.~D.,  2003, \mn@doi [\mnras] {10.1046/j.1365-8711.2003.06969.x},
  \href {https://ui.adsabs.harvard.edu/abs/2003MNRAS.345..575M} {345, 575}

\bibitem[\protect\citeauthoryear{{Matzner} \& {McKee}}{{Matzner} \&
  {McKee}}{1999}]{MM99}
{Matzner} C.~D.,  {McKee} C.~F.,  1999, \mn@doi [\apj] {10.1086/306571}, \href
  {https://ui.adsabs.harvard.edu/abs/1999ApJ...510..379M} {510, 379}

\bibitem[\protect\citeauthoryear{{M{\'e}sz{\'a}ros}}{{M{\'e}sz{\'a}ros}}{2006}]{Mesz06}
{M{\'e}sz{\'a}ros} P.,  2006, \mn@doi [Reports on Progress in Physics]
  {10.1088/0034-4885/69/8/R01}, \href
  {http://adsabs.harvard.edu/abs/2006RPPh...69.2259M} {69, 2259}

\bibitem[\protect\citeauthoryear{{Metzger} et~al.,}{{Metzger}
  et~al.}{2010}]{Metz10}
{Metzger} B.~D.,  et~al., 2010, \mn@doi [\mnras]
  {10.1111/j.1365-2966.2010.16864.x}, \href
  {https://ui.adsabs.harvard.edu/abs/2010MNRAS.406.2650M} {406, 2650}

\bibitem[\protect\citeauthoryear{{Mizuta} \& {Aloy}}{{Mizuta} \&
  {Aloy}}{2009}]{MA09}
{Mizuta} A.,  {Aloy} M.~A.,  2009, \mn@doi [\apj]
  {10.1088/0004-637X/699/2/1261}, \href
  {https://ui.adsabs.harvard.edu/abs/2009ApJ...699.1261M} {699, 1261}

\bibitem[\protect\citeauthoryear{{Mizuta}, {Yamasaki}, {Nagataki}  \&
  {Mineshige}}{{Mizuta} et~al.}{2006}]{Miz06}
{Mizuta} A.,  {Yamasaki} T.,  {Nagataki} S.,   {Mineshige} S.,  2006, \mn@doi
  [\apj] {10.1086/507861}, \href
  {https://ui.adsabs.harvard.edu/abs/2006ApJ...651..960M} {651, 960}

\bibitem[\protect\citeauthoryear{{Morsony}, {Lazzati}  \& {Begelman}}{{Morsony}
  et~al.}{2007}]{MLB07}
{Morsony} B.~J.,  {Lazzati} D.,   {Begelman} M.~C.,  2007, \mn@doi [\apj]
  {10.1086/519483}, \href
  {https://ui.adsabs.harvard.edu/abs/2007ApJ...665..569M} {665, 569}

\bibitem[\protect\citeauthoryear{{Nagakura}, {Hotokezaka}, {Sekiguchi},
  {Shibata}  \& {Ioka}}{{Nagakura} et~al.}{2014}]{Nag14}
{Nagakura} H.,  {Hotokezaka} K.,  {Sekiguchi} Y.,  {Shibata} M.,   {Ioka} K.,
  2014, \mn@doi [\apjl] {10.1088/2041-8205/784/2/L28}, \href
  {https://ui.adsabs.harvard.edu/abs/2014ApJ...784L..28N} {784, L28}

\bibitem[\protect\citeauthoryear{{Obergaulinger} \& {Aloy}}{{Obergaulinger} \&
  {Aloy}}{2019}]{OA19}
{Obergaulinger} M.,  {Aloy} M.~{\'A}.,  2019, arXiv e-prints, \href
  {https://ui.adsabs.harvard.edu/abs/2019arXiv190901105O} {}

\bibitem[\protect\citeauthoryear{{Perets}, {Li}, {Lombardi}  \&
  {Milcarek}}{{Perets} et~al.}{2016}]{Per16}
{Perets} H.~B.,  {Li} Z.,  {Lombardi} Jr. J.~C.,   {Milcarek} Jr. S.~R.,  2016,
  \mn@doi [\apj] {10.3847/0004-637X/823/2/113}, \href
  {https://ui.adsabs.harvard.edu/abs/2016ApJ...823..113P} {823, 113}

\bibitem[\protect\citeauthoryear{{Petrosian}, {Kitanidis}  \&
  {Kocevski}}{{Petrosian} et~al.}{2015}]{PKK15}
{Petrosian} V.,  {Kitanidis} E.,   {Kocevski} D.,  2015, \mn@doi [\apj]
  {10.1088/0004-637X/806/1/44}, \href
  {http://adsabs.harvard.edu/abs/2015ApJ...806...44P} {806, 44}

\bibitem[\protect\citeauthoryear{{Piran}}{{Piran}}{2004}]{pir04}
{Piran} T.,  2004, \mn@doi [Reviews of Modern Physics]
  {10.1103/RevModPhys.76.1143}, \href
  {http://adsabs.harvard.edu/abs/2004RvMP...76.1143P} {76, 1143}

\bibitem[\protect\citeauthoryear{Press, Teukolsky, Vetterling  \&
  Flannery}{Press et~al.}{2007}]{Press07}
Press W.~H.,  Teukolsky S.~A.,  Vetterling W.~T.,   Flannery B.~P.,  2007,
  Numerical Recipes 3rd Edition: The Art of Scientific Computing, 3 edn.
Cambridge University Press, USA

\bibitem[\protect\citeauthoryear{{Ramirez-Ruiz}, {Celotti}  \&
  {Rees}}{{Ramirez-Ruiz} et~al.}{2002}]{RR02}
{Ramirez-Ruiz} E.,  {Celotti} A.,   {Rees} M.~J.,  2002, \mn@doi [\mnras]
  {10.1046/j.1365-8711.2002.05995.x}, \href
  {https://ui.adsabs.harvard.edu/abs/2002MNRAS.337.1349R} {337, 1349}

\bibitem[\protect\citeauthoryear{{Rhoads}}{{Rhoads}}{1997}]{Rh97}
{Rhoads} J.~E.,  1997, \mn@doi [\apjl] {10.1086/310876}, \href
  {https://ui.adsabs.harvard.edu/abs/1997ApJ...487L...1R} {487, L1}

\bibitem[\protect\citeauthoryear{{Rhoads}}{{Rhoads}}{1999}]{Rhoads99}
{Rhoads} J.~E.,  1999, \mn@doi [\apj] {10.1086/307907}, \href
  {https://ui.adsabs.harvard.edu/abs/1999ApJ...525..737R} {525, 737}

\bibitem[\protect\citeauthoryear{{Rueda}, {Ruffini}  \& {Wang}}{{Rueda}
  et~al.}{2019}]{RRW19}
{Rueda} J.~A.,  {Ruffini} R.,   {Wang} Y.,  2019, \mn@doi [Universe]
  {10.3390/universe5050110}, \href
  {https://ui.adsabs.harvard.edu/abs/2019Univ....5..110R} {5, 110}

\bibitem[\protect\citeauthoryear{{Suwa} \& {Ioka}}{{Suwa} \&
  {Ioka}}{2011}]{SI11}
{Suwa} Y.,  {Ioka} K.,  2011, \mn@doi [\apj] {10.1088/0004-637X/726/2/107},
  \href {https://ui.adsabs.harvard.edu/abs/2011ApJ...726..107S} {726, 107}

\bibitem[\protect\citeauthoryear{{Tanvir}, {Levan}, {Fruchter}, {Hjorth},
  {Hounsell}, {Wiersema}  \& {Tunnicliffe}}{{Tanvir} et~al.}{2013}]{TL13}
{Tanvir} N.~R.,  {Levan} A.~J.,  {Fruchter} A.~S.,  {Hjorth} J.,  {Hounsell}
  R.~A.,  {Wiersema} K.,   {Tunnicliffe} R.~L.,  2013, \mn@doi [\nat]
  {10.1038/nature12505}, \href
  {https://ui.adsabs.harvard.edu/abs/2013Natur.500..547T} {500, 547}

\bibitem[\protect\citeauthoryear{{Tchekhovskoy}, {Narayan}  \&
  {McKinney}}{{Tchekhovskoy} et~al.}{2010}]{Tch10}
{Tchekhovskoy} A.,  {Narayan} R.,   {McKinney} J.~C.,  2010, \mn@doi [\na]
  {10.1016/j.newast.2010.03.001}, \href
  {https://ui.adsabs.harvard.edu/abs/2010NewA...15..749T} {15, 749}

\bibitem[\protect\citeauthoryear{{Troja} et~al.,}{{Troja} et~al.}{2018}]{Tro18}
{Troja} E.,  et~al., 2018, \mn@doi [Nature Communications]
  {10.1038/s41467-018-06558-7}, \href
  {https://ui.adsabs.harvard.edu/abs/2018NatCo...9.4089T} {9, 4089}

\bibitem[\protect\citeauthoryear{{Tsvetkova} et~al.,}{{Tsvetkova}
  et~al.}{2017}]{TSv17}
{Tsvetkova} A.,  et~al., 2017, \mn@doi [\apj] {10.3847/1538-4357/aa96af}, \href
  {https://ui.adsabs.harvard.edu/abs/2017ApJ...850..161T} {850, 161}

\bibitem[\protect\citeauthoryear{{Vink} \& {de Koter}}{{Vink} \& {de
  Koter}}{2005}]{VdK05}
{Vink} J.~S.,  {de Koter} A.,  2005, \mn@doi [\aap]
  {10.1051/0004-6361:20052862}, \href
  {https://ui.adsabs.harvard.edu/abs/2005A%26A...442..587V} {442, 587}

\bibitem[\protect\citeauthoryear{{Vink}, {de Koter}  \& {Lamers}}{{Vink}
  et~al.}{2001}]{Vink01}
{Vink} J.~S.,  {de Koter} A.,   {Lamers} H.~J.~G.~L.~M.,  2001, \mn@doi [\aap]
  {10.1051/0004-6361:20010127}, \href
  {https://ui.adsabs.harvard.edu/abs/2001A%26A...369..574V} {369, 574}

\bibitem[\protect\citeauthoryear{{Wang} \& {Dai}}{{Wang} \& {Dai}}{2011}]{WD11}
{Wang} F.~Y.,  {Dai} Z.~G.,  2011, \mn@doi [\apjl]
  {10.1088/2041-8205/727/2/L34}, \href
  {https://ui.adsabs.harvard.edu/abs/2011ApJ...727L..34W} {727, L34}

\bibitem[\protect\citeauthoryear{{Wang}, {Zou}, {Liu}, {Liao}, {Liu}, {Chai}
  \& {Xia}}{{Wang} et~al.}{2019}]{Wang2019}
{Wang} F.,  {Zou} Y.-C.,  {Liu} F.,  {Liao} B.,  {Liu} Y.,  {Chai} Y.,   {Xia}
  L.,  2019, arXiv e-prints, \href
  {http://adsabs.harvard.edu/abs/2019arXiv190205489W} {}

\bibitem[\protect\citeauthoryear{{Wei} \& {Gao}}{{Wei} \& {Gao}}{2003}]{WG03}
{Wei} D.~M.,  {Gao} W.~H.,  2003, \mn@doi [\mnras]
  {10.1046/j.1365-8711.2003.06971.x}, \href
  {https://ui.adsabs.harvard.edu/abs/2003MNRAS.345..743W} {345, 743}

\bibitem[\protect\citeauthoryear{{Woosley}}{{Woosley}}{1993}]{Woos93}
{Woosley} S.~E.,  1993, \mn@doi [\apj] {10.1086/172359}, \href
  {http://adsabs.harvard.edu/abs/1993ApJ...405..273W} {405, 273}

\bibitem[\protect\citeauthoryear{Woosley \& Bloom}{Woosley \&
  Bloom}{2006}]{WB06}
Woosley S.,  Bloom J.,  2006, Annu. Rev. Astron. Astrophys., 44, 507

\bibitem[\protect\citeauthoryear{{Woosley} \& {Heger}}{{Woosley} \&
  {Heger}}{2006}]{WH06}
{Woosley} S.~E.,  {Heger} A.,  2006, \mn@doi [\apj] {10.1086/498500}, \href
  {https://ui.adsabs.harvard.edu/abs/2006ApJ...637..914W} {637, 914}

\bibitem[\protect\citeauthoryear{{Xue}, {Zhang}  \& {Zhu}}{{Xue}
  et~al.}{2019}]{Xue19}
{Xue} L.,  {Zhang} F.-W.,   {Zhu} S.-Y.,  2019, arXiv e-prints, \href
  {http://adsabs.harvard.edu/abs/2019arXiv190407767X} {}

\bibitem[\protect\citeauthoryear{{Yonetoku}, {Murakami}, {Nakamura},
  {Yamazaki}, {Inoue}  \& {Ioka}}{{Yonetoku} et~al.}{2004}]{Yon04}
{Yonetoku} D.,  {Murakami} T.,  {Nakamura} T.,  {Yamazaki} R.,  {Inoue} A.~K.,
   {Ioka} K.,  2004, \mn@doi [\apj] {10.1086/421285}, \href
  {http://adsabs.harvard.edu/abs/2004ApJ...609..935Y} {609, 935}

\bibitem[\protect\citeauthoryear{{Yonetoku}, {Yamazaki}, {Nakamura}  \&
  {Murakami}}{{Yonetoku} et~al.}{2005}]{Yon05}
{Yonetoku} D.,  {Yamazaki} R.,  {Nakamura} T.,   {Murakami} T.,  2005, \mn@doi
  [\mnras] {10.1111/j.1365-2966.2005.09398.x}, \href
  {https://ui.adsabs.harvard.edu/abs/2005MNRAS.362.1114Y} {362, 1114}

\bibitem[\protect\citeauthoryear{{Yoon}, {Woosley}  \& {Langer}}{{Yoon}
  et~al.}{2010}]{Yoon10}
{Yoon} S.-C.,  {Woosley} S.~E.,   {Langer} N.,  2010, \mn@doi [\apj]
  {10.1088/0004-637X/725/1/940}, \href
  {https://ui.adsabs.harvard.edu/abs/2010ApJ...725..940Y} {725, 940}

\bibitem[\protect\citeauthoryear{{Yoon}, {Dierks}  \& {Langer}}{{Yoon}
  et~al.}{2012}]{Yoon12}
{Yoon} S.-C.,  {Dierks} A.,   {Langer} N.,  2012, \mn@doi [\aap]
  {10.1051/0004-6361/201117769}, \href
  {https://ui.adsabs.harvard.edu/abs/2012A%26A...542A.113Y} {542, A113}

\bibitem[\protect\citeauthoryear{{Yu}, {Wang}, {Dai}  \& {Cheng}}{{Yu}
  et~al.}{2015}]{Yu15}
{Yu} H.,  {Wang} F.~Y.,  {Dai} Z.~G.,   {Cheng} K.~S.,  2015, \mn@doi [\apjs]
  {10.1088/0067-0049/218/1/13}, \href
  {http://adsabs.harvard.edu/abs/2015ApJS..218...13Y} {218, 13}

\bibitem[\protect\citeauthoryear{{Y{\"u}ksel}, {Kistler}, {Beacom}  \&
  {Hopkins}}{{Y{\"u}ksel} et~al.}{2008}]{Yuk08}
{Y{\"u}ksel} H.,  {Kistler} M.~D.,  {Beacom} J.~F.,   {Hopkins} A.~M.,  2008,
  \mn@doi [\apjl] {10.1086/591449}, \href
  {https://ui.adsabs.harvard.edu/abs/2008ApJ...683L...5Y} {683, L5}

\bibitem[\protect\citeauthoryear{{Zhang} \& {Fryer}}{{Zhang} \&
  {Fryer}}{2001}]{ZF01}
{Zhang} W.,  {Fryer} C.~L.,  2001, \mn@doi [\apj] {10.1086/319734}, \href
  {https://ui.adsabs.harvard.edu/abs/2001ApJ...550..357Z} {550, 357}

\bibitem[\protect\citeauthoryear{{Zhang} \& {M{\'e}sz{\'a}ros}}{{Zhang} \&
  {M{\'e}sz{\'a}ros}}{2004}]{ZM04}
{Zhang} B.,  {M{\'e}sz{\'a}ros} P.,  2004, \mn@doi [International Journal of
  Modern Physics A] {10.1142/S0217751X0401746X}, \href
  {http://adsabs.harvard.edu/abs/2004IJMPA..19.2385Z} {19, 2385}

\bibitem[\protect\citeauthoryear{{Zhang}, {Woosley}  \& {MacFadyen}}{{Zhang}
  et~al.}{2003}]{ZWM03}
{Zhang} W.,  {Woosley} S.~E.,   {MacFadyen} A.~I.,  2003, \mn@doi [\apj]
  {10.1086/367609}, \href
  {https://ui.adsabs.harvard.edu/abs/2003ApJ...586..356Z} {586, 356}

\bibitem[\protect\citeauthoryear{{de Mink}, {Langer}, {Izzard}, {Sana}  \& {de
  Koter}}{{de Mink} et~al.}{2013}]{deMink13}
{de Mink} S.~E.,  {Langer} N.,  {Izzard} R.~G.,  {Sana} H.,   {de Koter} A.,
  2013, \mn@doi [\apj] {10.1088/0004-637X/764/2/166}, \href
  {https://ui.adsabs.harvard.edu/abs/2013ApJ...764..166D} {764, 166}

\bibitem[\protect\citeauthoryear{{van Dokkum} \& {van der Marel}}{{van Dokkum}
  \& {van der Marel}}{2007}]{vD07}
{van Dokkum} P.~G.,  {van der Marel} R.~P.,  2007, \mn@doi [\apj]
  {10.1086/509633}, \href
  {https://ui.adsabs.harvard.edu/abs/2007ApJ...655...30V} {655, 30}

\makeatother
\end{thebibliography}

%%%%%%%%%%%%%

%%%%%%%%%%%%%%%%% APPENDICES %%%%%%%%%%%%%%%%%%%%%

%%%%%%%%%%%%%%%%%%%%%%%%%%%%%%%%%%%%%%%%%%%%%%%%%%

% Don't change these lines
\bsp	% typesetting comment
\label{lastpage}
\end{document}